\documentclass[pra,aps,amssymb,twocolumn,noshowpacs]{revtex4-1}
\usepackage{color}
\usepackage{graphicx}
\usepackage{hyperref}
\usepackage{amsmath}
\usepackage{latexsym}
\usepackage{amssymb}
\usepackage{mathrsfs}
\usepackage{mathtools}
\usepackage{layout}
\usepackage{verbatim}
\usepackage{multirow}
\usepackage{bm}
\usepackage{amsfonts,epsfig}

\begin{document}

\title{Doppler-resilient ground-Rydberg transition and its application in high-fidelity entangling gates with neutral atoms}
\date{\today}
\author{Xiao-Feng Shi}
\affiliation{School of Physics and Optoelectronic Engineering, Xidian University, Xi'an 710071, China}

\begin{abstract}
  The motion-induced dephasing is a severe problem that limits the accuracy of a quantum control process by using external laser fields in neutral Rydberg atoms. This dephasing is a major issue that limits the realizable fidelity of a quantum entangling gate with neutral atoms when there is a {\it gap} time for the Rydberg atom to drift freely. We find that such a dephasing can be largely suppressed by using a transition in a `V'-type dual-rail configuration. The left~(right) arm of this `V' represents a transition to a Rydberg state $|r_{1(2)}\rangle$ with a Rabi frequency $\Omega e^{ikz}~(\Omega e^{-ikz})$, where $z$ is frozen without atomic drift, but changes linearly in each experimental cycle. Such a configuration is equivalent to a transition between the ground state and a hybrid and time-dependent Rydberg state with a Rabi frequency $\sqrt2\Omega$, such that there is no phase error whenever the state returns to the ground state. We study two applications of this method. First, it is possible to faithfully transfer the atomic state between a hyperfine ground state $|1\rangle$ and Rydberg states $|r_{1(2)}\rangle$ with no {\it gap} time between the excitation and deexcitation. Second, by adding infrared laser fields to induce transition between $|r_{1(2)}\rangle$ and a nearby Rydberg state $|r_3\rangle$ via a largely detuned low-lying intermediate state in the {\it gap} time, the atom can keep its internal state in the Rydberg level as well as adjust the population branching in $|r_{1(2)}\rangle$ during the {\it gap} time. This allows an almost perfect Rydberg deexcitation after the {\it gap} time, making it possible to recover a high fidelity in the Rydberg blockade gate. The theory paves the way for high-fidelity quantum control over neutral Rydberg atoms without cooling qubits to the motional ground states in optical traps.

\end{abstract}
\maketitle

\section{introduction}\label{sec01}
Inducing a ground-Rydberg state transfer via external laser fields in ultracold neutral atoms is prevalent in the investigation of quantum computing~\cite{Saffman2010,Saffman2016,Weiss2017}, single-photon quantum optics~\cite{Dudin2012,Peyronel2012,Firstenberg2013,Li2013,Gorniaczyk2014,Tiarks2014,Busche2017,Lampen2018,Adams2019}, and quantum many-body physics~\cite{Weimer2009}. For laser-cooled neutral atoms, the achievable precision of these tasks is subjected to a number of issues, among which is the well-known motion-induced Doppler dephasing. The different phases of the laser fields seen by the drifting atom along the time hamper the state transfer, leading to errors in both the population and the phase. For the design of a neutral-atom entangling gate~\cite{PhysRevLett.85.2208}, both the population and the phase of the transferred state need to be accurate; this is why the Doppler dephasing is a critical issue~\cite{Wilk2010,Saffman2011,Saffman2016,Levine2019,Graham2019}. In principle, excitation of an atom by two counter-propagating light waves that form a standing wave can eliminate the Doppler effect. But this method leads to a space-dependent Rabi frequency because of the sinusoidal modulation of the field amplitude in a standing wave, and thus it is not of practical use for achieving high fidelity in quantum control over flying atoms. It was shown in Ref.~\cite{Ryabtsev2011} that a three-photon transition via focusing three beams of different lasers at a common spot can lead to a Doppler-free transition between ground and Rydberg states of a rubidium atom at that spot. Nevertheless, the necessary condition of largely detuning the two intermediate states in the transition chain caps the achievable magnitude of the effective Rabi frequency~\cite{Ryabtsev2011}. This raises the question of whether there is any practical route to eliminate Doppler dephasing in the two-photon excitation of Rydberg states that is widely used in experiments~\cite{Wilk2010,Isenhower2010,Zhang2010,Maller2015,Zeng2017,Levine2018,Picken2018,Levine2019,Graham2019}.

\begin{figure}
\includegraphics[width=3.2in]
{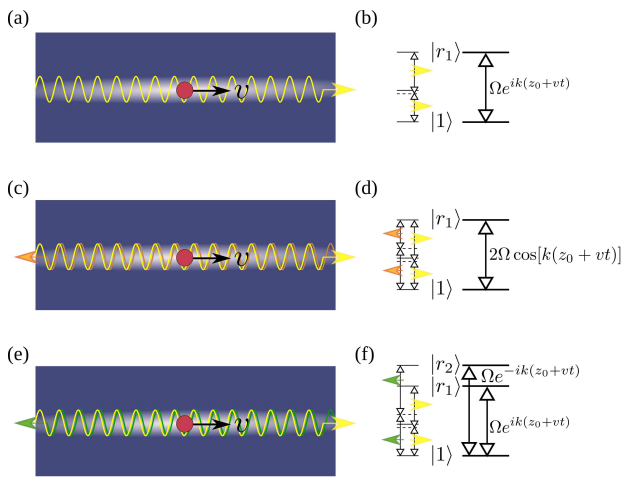}
 \caption{(a) An atom is optically pumped by laser fields traveling along $\mathbf{z}$ when it is drifting in free space whose velocity along $\mathbf{z}$ is $v$. (b) Due to the atomic drift, the Rabi frequency is $\Omega e^{ik(z_0+vt)}$, where $z_0$ is the initial coordinate of the atom, $k$ is the wavevector, and $t$ is the time. For a Rabi transition between the ground state $|1\rangle$ and an even-parity Rydberg state $|r_1\rangle$, $|1\rangle\leftrightarrow|r_1\rangle$ is a resonant two-photon transition via a largely detuned intermediate state. (c) By adding another set of laser fields that propagate with directions that are opposite to those in (a), the effective Rabi frequency becomes $\Omega e^{ik(z_0+vt)}+\Omega e^{-ik(z_0+vt)}= 2\Omega \cos[k(z_0+vt)]$, shown in (d). (e) and (f) show an alternative dephasing-resilient protocol, where one two-photon Rabi frequency is used for addressing each of the two Rydberg states. \label{figure01} }
\end{figure}

In this work, we demonstrate suppression of Doppler dephasing in a two-photon excitation and deexcitation of Rydberg states. Our theory takes advantage of quantum interference between two transitions from a common ground state, $|1\rangle$, to two different Rydberg eigenstates, $|r_{1(2)}\rangle$. We show two methods to suppress this dephasing, where the latter is simpler to implement and will be used in most numerical simulations later, but the former is easier to understand although the two methods originate from the same pattern of quantum interference. We start with the first method. The Doppler dephasing in the transition $|1\rangle\leftrightarrow|r_1\rangle$ originates from the constant phase change of the Rabi frequency during the optical pumping shown in Fig.~\ref{figure01}(a), where $v$ is the atomic velocity along the propagation direction of laser light. To suppress the dephasing, we turn on another set of laser fields~[shown in Fig.~\ref{figure01}(b)] that propagate in opposite direction to those in Fig.~\ref{figure01}(a). This introduces a new two-photon Rabi frequency $\Omega e^{-ik(z_0+vt)}$~(or $-\Omega e^{-ik(z_0+vt)}$) such that $|1\rangle\leftrightarrow|r_1\rangle$ is characterized by a Rabi frequency $2\Omega \cos[k(z_0+vt)]$~(or $2i\Omega \sin[k(z_0+vt)]$). In this method, an atom initialized in $|1\rangle$ can be simultaneously excited to $|r_1\rangle$ and $|r_2\rangle$ by the Rabi frequencies $2\Omega \cos[k(z_0+vt)]$ and $2i\Omega \sin[k(z_0+vt)]$, respectively; then, the three-level system formed by $|1\rangle,~|r_1\rangle$, and $|r_2\rangle$ undergoes the following transition
\begin{eqnarray}
 |1\rangle\xleftrightarrow[\text{Effective}]{2\Omega} w_1(t)|r_1\rangle+w_2(t)|r_2\rangle. \label{sec01eq01}
\end{eqnarray}
For $kv\ll \Omega$, we find 
\begin{eqnarray}
  w_\alpha(t) &=&i^{2-\alpha} \Omega\sum_{\eta=\pm}\frac{f_\alpha (kz_0) - f_\alpha[kz_0+(kv+\eta \Omega)t]}{2(kv+\eta \Omega)} , \nonumber\\\label{w1w2}
\end{eqnarray}
where $(f_1,~f_2)=(\sin,~\cos)$ and $\alpha=1$ and 2. Remarkably, $\sum_\alpha|w_\alpha(t)|^2$ becomes $1~(0)$ at $t=\frac{m\pi}{2\Omega}$ when $m$ is an odd~(even) integer in Eq.~(\ref{w1w2}). Quantum control over Rydberg atoms often requires the population to oscillate fully between the ground and Rydberg states; this is achieved by Eqs.~(\ref{sec01eq01}) and~(\ref{w1w2}). More importantly, there is no phase error for the ground state during the quantum evolution of Eqs.~(\ref{sec01eq01}) and~(\ref{w1w2}), indicating a rigorous suppression of Doppler dephasing.

Although the above method can eliminate the Doppler dephasing, it requires four sets of laser fields to work. A variant of the above method by using two sets of laser fields is shown in Fig.~\ref{figure01}(c), where $|1\rangle$ is connected with $|r_1\rangle$ and $|r_2\rangle$ by a Rabi frequency of $\Omega e^{ik(z_0+vt)}$ and $\Omega e^{-ik(z_0+vt)}$, respectively. By defining $|r_\pm\rangle=(|r_1\rangle\pm|r_2\rangle)/\sqrt2$, we have $|1\rangle\xleftrightarrow[\text{Effective}]{\sqrt2\Omega} \sum_\pm w_\pm(t)|r_\pm\rangle$, where $w_\pm(t)$ is defined in a similar way as $w_{1(2)}$ in Eq.~(\ref{w1w2}). This latter method is simpler in that it only needs one two-photon transition to address either $|r_{1}\rangle$ or $|r_{2}\rangle$, while the method in Eq.~(\ref{w1w2}) needs two. Although this latter method should, in principle, suffer from the recoil problem, its effect is negligible as shown later.

Among a number of issues~\cite{DeLeseleuc2018,Levine2018} in the study of Rydberg blockade gate~\cite{PhysRevLett.85.2208}, Doppler dephasing is a critical issue limiting its fidelity~\cite{Wilk2010,Saffman2011,Saffman2016}. By using either of the two methods described above, there is no phase error in the ground state. The total population in the two Rydberg states can reach 1 at  $t=\frac{\pi}{\sqrt2\Omega}$~[in the case of Fig.~\ref{figure01}(f)] with negligible error. To bring the Rydberg state back to the ground state, one can use infrared laser fields to transfer population between Rydberg states $|r_{1(2)}\rangle$ and another Rydberg state $|r_{3}\rangle$ via a low-lying intermediate state. The infrared pumping lets the Rydberg atom adjust its population branching between $|r_{1}\rangle$ and $|r_{2}\rangle$, which prepares for an almost perfect state restoration after the gap time. As shown later, these three properties lead to a high fidelity for the Rydberg blockade gate with a gap time~\cite{PhysRevLett.85.2208}. For Rydberg gates realized with no gap time~\cite{Shi2018Accuv1,Graham2019}, our theory can also be used with the stage of infrared pumping removed.

The remainder of this work is organized as follows. In Sec.~\ref{sec02}, we analyze in detail the two methods mentioned above and show the suppression of the Doppler dephasing in the ground-Rydberg transition via two-photon transitions when there is no gap time. In Sec.~\ref{sec03}, we first study conditions to restore the state from Rydberg to ground state during the gap time when the Rydberg atom is in free flight. We then study the achievable gate fidelity of the controlled-Z gate based on Rydberg blockade in our theory. In Sec.~\ref{sec04}, we discuss the application of our theory to other types of quantum gates and compare the theory in this work with that in~\cite{Shi2019Dopp2}. Section~\ref{sec05} summarizes the work.

\begin{figure*}[ht]
\includegraphics[width=7.0in]
{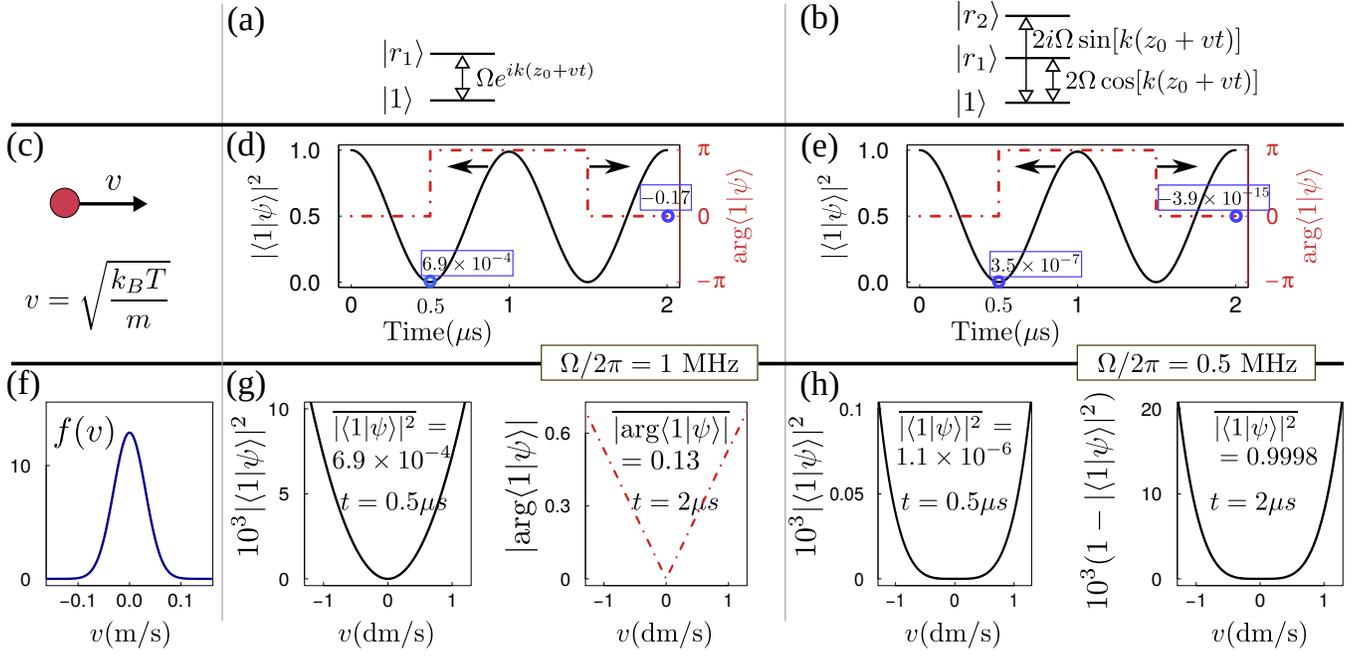}
 \caption{(a) A two-photon resonant transition $|1\rangle\leftrightarrow|r_1\rangle$ with a Rabi frequency $\Omega e^{ik(z_0+vt)}$. (b) A transition chain $|r_1\rangle\leftrightarrow|1\rangle\leftrightarrow |r_2\rangle$ with Rabi frequencies $2\Omega \cos{[k(z_0+vt)]}$ and $2i\Omega \sin{[k(z_0+vt)]}$. (c) An atom drifts with a velocity whose $\mathbf{z}$-component is $v$, where $v=\sqrt{k_BT/m}=3.1$cm/s when $T=10~\mu$K for rubidium-87. (d) and (e) show the population and phase dynamics of the ground-state component $|1\rangle$ under the external drive in (a) and (b) respectively. The initial state is $|\psi(0)\rangle=|1\rangle$ and the drift is shown in (c) with $z_0=0$. $\Omega/2\pi=1$ and $0.5$~MHz in (d) and (e), respectively. The population in $|1\rangle$ at $t=0.5\mu$s are $6.94\times10^{-4}$ and $3.54\times10^{-7}$ in (d) and (e), respectively, while the phase errors for $|1\rangle$ at $t=2\mu$s are $0.17$ and $<10^{-10}$ in (d) and (e), respectively. For different $z_0$, the solid curves (population) in (d) and (e) stay the same, while the dashed curves~(phase) change, but the values of $\langle1|\psi\rangle$ at $t=0.5$ and $2~\mu$s are the same for both cases. (f) Maxwell distribution of the atomic velocity along $\mathbf{z}$ at $T=10~\mu$K. (g) $|\langle1|\psi(0.5\mu s)\rangle|^2$ scaled up by $10^3$~(left), and $|$arg$\langle1|\psi(2\mu s)\rangle|$~(right) as a function of $v$ with the external drive in (a). The average errors of the population and phase are $6.93\times10^{-4}$ and $0.13$ in the left and right subfigures in (g), respectively, and the average population error in $|1\rangle$ at $2~\mu$s is $1.42\times10^{-5}$ (not shown here). (h) The left and right panel show $|\langle1|\psi(0.5\mu s)\rangle|^2$ and $(1-|\langle1|\psi(2\mu s)\rangle|^2)$ scaled up by $10^3$, respectively, as a function of $v$ with the external drive in (b). The populations averaged over the Maxwell distribution are $1.1\times10^{-6}$ and $0.9998$ at $0.5$ and $2~\mu$s, respectively. The phase error at $2~\mu$s is below $10^{-10}$~(not shown here). Figures (g) and (h) stay the same for different $z_0$. Here $k =k_-$ and $k_\pm\equiv 2\pi(1/474\pm1/795)$nm$^{-1}$ in this work.  \label{figure02} }
\end{figure*}

\section{Case 1: When there is no gap time}\label{sec02}
In this section, we show that for atoms cooled only to about $10~\mu$K and without any wait period for them to remain in the Rydberg shelving states during free flight, it is possible to excite an atom from ground to Rydberg states and restore the state back to the ground state with negligible error. In Sec.~\ref{sec02a}, we consider a model with four sets of Rabi transitions between the ground state and Rydberg states. The more useful method realized only by two sets of Rabi transitions is then presented in Sec.~\ref{sec02b}.

\subsection{A method with four sets of laser fields}\label{sec02a}

The Doppler-dephasing-insensitive state transfer in a dual-rail counter-rotating Rabi transition can be understood in three steps.\newline\newline
(i) First, although an atom can drift with a speed that is nonzero in all the three directions, $\mathbf{x},~\mathbf{y}$, and $\mathbf{z}$, the Rabi frequency characterizing the state transfer from $|1\rangle$ to a Rydberg state $|r_1\rangle$ is only influenced by the drift along the propagation direction of the laser fields, shown in Fig.~\ref{figure01}(a). If $|1\rangle\leftrightarrow|r_1\rangle$ arises under the excitation of lasers that travel along $\mathbf{z}$, the Rabi frequency is $\Omega e^{ik(z_0+vt)}$, where $z_0$ is the coordinate of the atom at $t=0$, $k$ is the wavevector of the laser fields, and $v$ is the atomic speed along $\mathbf{z}$, shown in Fig.~\ref{figure01}(b). The drift along $\mathbf{x}$ and $\mathbf{y}$ may result in change of the magnitude of $\Omega$~\cite{Graham2019}, but this problem is avoided when the waist of laser beam is large enough compared with the size of the dipole trap and the atomic drift as discussed in~\cite{Shi2018pra_m,Shi2018Accuv1,Covey2019prappl}.\newline\newline
(ii) Second, when $|1\rangle\leftrightarrow|r_1\rangle$ is a two-photon transition, Figs.~\ref{figure01}(c) and~\ref{figure01}(d) show that it is possible to use another set of laser fields that propagate oppositely to those in Fig.~\ref{figure01}(a) to introduce a new two-photon Rabi frequency $\Omega e^{-ik(z_0+vt)}$. This induces a resonant two-photon transition via an intermediate state. If only one intermediate state is used, the four laser fields in Fig.~\ref{figure01}(d) can all couple to it, but the single-photon detuning for $\Omega e^{ik(z_0+vt)}$ and that for $\Omega e^{-ik(z_0+vt)}$ should have a difference that is large compared with $\Omega$. Then, $|1\rangle\leftrightarrow|r_1\rangle$ is driven by a Rabi frequency $2\Omega\cos[k(z_0+vt)]$. The appearance of this Rabi frequency mimics the well known phenomenon that the action of two symmetrically detuned laser pulses is equivalent to that of a monochromatic field whose amplitude is sinusoidally modulated in time~\cite{Goreslavsky1980}. \newline\newline
(iii) Third, the Rabi frequency $2\Omega\cos[k(z_0+vt)]$ can lead to a transition from $|1\rangle$ to $|r_1\rangle$, but with a $z_0$- and $v$-dependent time $t_\pi$ that is determined by $|\sin(z_0+kvt_\pi)-\sin z_0|=k|v|\pi/(2\Omega)$. However, this method of determining $t_\pi$ is valid only if $z_0$ and $v$ are both known. For a neutral atom trapped in an optical trap before each experimental cycle, the values of $z_0$ and $v$ in different experimental runs can differ because of the thermal distribution. Thus, the setup in Fig.~\ref{figure01}(d) is not useful for pumping the state from $|1\rangle$ to $|r_1\rangle$. However, we can add a new pair of two-photon Rabi transitions from $|1\rangle$ to another state $|r_2\rangle$. For two transitions with Rabi frequencies $\Omega e^{ik(z_0+vt)}$ and $-\Omega e^{-ik(z_0+vt)}$, $|1\rangle\leftrightarrow|r_2\rangle$ is characterized by a Rabi frequency $2i\Omega\sin[k(z_0+vt)]$. As a result, for an initial state in $|1\rangle$, the three-level system formed by $|1\rangle,~|r_1\rangle$, and $|r_2\rangle$ follows the transition shown in Eq.~(\ref{sec01eq01}). If the intermediate state for $|1\rangle\leftrightarrow|r_{1(2)}\rangle$ is the $5P_{1/2}$ state of rubidium, $k$ is $2\pi(1/474-1/795)$nm$^{-1}$ for two counter-propagating laser fields. Then, $kv\approx0.17$~rad$/\mu s$ with a typical velocity of $v=v_{\text{rms}}\equiv\sqrt{k_BT/m}=3.1$cm/s when $T=10~\mu$K. For $kv\ll \Omega$, the depopulation of $|1\rangle$ occurs as follows: a transition with a Rabi frequency $2\Omega$ takes the population from $|1\rangle$ to $w_1(t)|r_1\rangle+w_2(t)|r_2\rangle$, and during each small time $\delta t$, the change of wavefunction follows the rule of $[w_1(t+\delta t)-w_1(t)]~:~[w_2(t+\delta t)-w_2(t)]=\cos[k(z_0+vt)]~:~i\sin[k(z_0+vt)]$. This leads to Eq.~(\ref{w1w2}) and $|w_1(t)|^2+|w_2(t)|^2=\sin^2(\Omega t)$. We have numerically simulated this with random values of $z_0$ and found negligible error for $v_{\text{rms}}$ corresponding to $T$ up to $10~\mu$K. An interesting observation is that the mismatch between Eq.~(\ref{w1w2}) and the simulation data shrinks quickly when $z_0$ deviates from $0$. Although the population branching between the two upper states differ for different $z_0$ and $v$, the depopulation process of $|1\rangle$ follows the same pattern, demonstrating a rigorous suppression of Doppler dephasing.

Figure~\ref{figure02} shows the suppression of the Doppler dephasing for the above theory. For $z_0=0$, $v_{\text{rms}}$ for $T=10~\mu$K, and $\Omega/2\pi = 0.5$~MHz, the population error in $|1\rangle$ at $t=0.5~\mu$s is $3.54\times10^{-7}$. At $t=1~\mu$s, the population in $|1\rangle$ is only $0.989$; but it reaches $0.99992$ at $t=2~\mu$s, and the phase of $\langle 1|\psi(2\mu s)\rangle$ is negligible~(smaller than $10^{-10}$ according to the simulation data). These results also hold for any other $z_0$. When averaged over the Maxwell distribution~[see Fig.~\ref{figure02}(f)] of the atomic velocity, the populations in $|1\rangle$ are $10^{-6}$ and $0.9998$ at $t=0.5$ and $2~\mu$s, respectively, as shown in Fig.~\ref{figure02}(h). In comparison, the method for one Rabi transition with $\Omega/2\pi = 1$~MHz is also shown in Figs.~\ref{figure02}(a),~(d), and~(g). The population error at $t=0.5$~$\mu$s is $7\times10^{-4}$ in Fig.~\ref{figure02}(d), larger than that in Fig.~\ref{figure02}(e) by about 600 times. Although the population error in $|1\rangle$ is about $10^{-5}$ at $2~\mu$s, its phase error is larger than $0.1$~rad, either for a typical velocity $v_{\text{rms}}$ or averaged over the Maxwell distribution, shown in Figs.~\ref{figure02}(d) and~\ref{figure02}(g); similar error exists at $t=1~\mu$s. For our method shown in Figs.~\ref{figure02}(b),~\ref{figure02}(e),~and~\ref{figure02}(h), the population error in $|1\rangle$ with two Rabi cycles~(at $2~\mu$s) is much smaller than that with one Rabi cycle~(at $1~\mu$s). This is true for other values of $z_0$ and $T$, and is because the actual dynamics is slightly different from that described by Eq.~(\ref{w1w2}); the smaller $T$ is, the less prominent will this effect be. For this reason, it is necessary to use a $3\pi$ pulse to restore the state back to the ground state by our method.

\subsection{A method with two sets of laser fields}\label{sec02b}
As discussed in Sec.~\ref{sec01}, an alternative method is to use one Rabi frequency for each of the two Rydberg states, $|r_1\rangle$ and $|r_2\rangle$, as shown in Fig.~\ref{figure01}(f). Because the system dynamics in this method (with respect to a new basis $|r_\pm\rangle=(|r_1\rangle\pm|r_2\rangle)/\sqrt2$) is identical to the one shown in Fig.~\ref{figure02}, the population and phase dynamics in Fig.~\ref{figure02} will not be repeated here. Since this latter method is relatively simple and is easier to implement in experiments, we will give details about how to use it for the study of Rydberg blockade gate~\cite{PhysRevLett.85.2208}.

In Fig.~\ref{figure02}(e), the population error in $|1\rangle$ at the end of the two Rabi cycles is on the order of $10^{-4}$. Below, we show that this error can be suppressed by choosing an optimal Rabi frequency for the state restoration. The optimal condition is numerically analyzed based on a linear form of the phase accumulation in the presence of the atomic motion, as detailed below.

\subsubsection{ Linear phase accumulation }
The Hamiltonian for the system in Fig.~\ref{figure01}(f) can be written as
\begin{eqnarray}
  \hat{H}(t) &=&\frac{ \Omega }{2}\left(
  \begin{array}{ccc}
    0 & 0 & e^{-ik(z_0 + vt)}\\
    0 & 0 &e^{ik(z_0 + vt)}\\
    e^{ik(z_0 + vt)} &  e^{-ik(z_0 + vt)}&0
    \end{array}
  \right) \label{Hamil01}
\end{eqnarray}
in the basis $\{|r_2\rangle,|r_1\rangle,|1\rangle \}$. In the above Hamiltonian, we have ignored the velocity change of the atom upon absorption of the photon; the reason is given later. As an example, we consider a two-photon excitation of high-lying Rydberg states with $k_\pm =2\pi(1/474\pm1/795)$nm$^{-1}$~\cite{Baur2014,Tiarks2014}.

\begin{figure}
\includegraphics[width=3.2in]
{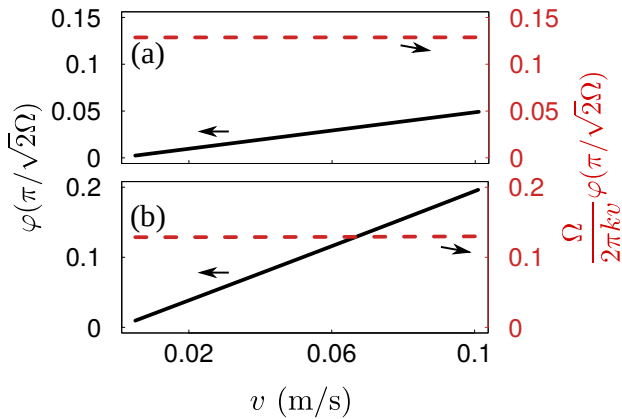}
\caption{The solid curve shows $\varphi(\pi/\sqrt2\Omega)$, defined in Eq.~(\ref{varphi01}), and the dashed curve shows the ratio between $\varphi$ and $\frac{2\pi kv}{\Omega}$, where $\Omega/2\pi=\sqrt2$~MHz. The wavevector $k$ is $k_-=2\pi(1/474-1/795)$nm$^{-1}$ in (a), and $k_+=2\pi(1/474+1/795)$nm$^{-1}$ in (b). For rubidium-87, the velocity $v$ is given by $v=\sqrt{k_BT/m}\in\{0.005,~0.1\}$m/s when $T\in\{0.3,~105\}~\mu$K. \label{figure003} }
\end{figure}

For a state initialized in the ground state, the time evolution of the system wavefunction $|\psi(t)\rangle = C_{\text{g}}(t)|1\rangle+C_{\text{r1}}(t)|r_1\rangle+C_{\text{r2}}(t)|r_2\rangle$ is captured by the time-ordering operator $U(t)\equiv\mathcal{T} e^{-i\int_0^t \hat{H}(\eta)d\eta}$, where we should bear in mind that $U(t)$ at a given time is also a function of $z_0,~v$, and $\Omega$. Because the direct effect of Doppler dephasing is the phase change of the Rabi frequency, the atomic drift results in phase modulation in both $C_{\text{r1}}(t)$ and $C_{\text{r2}}(t)$. The phase factor $\pm kz_0$ in the Rabi frequency, determined by the initial location of the qubit, is constant along the pumping process, and it is exactly compensated when the population returns to the ground state. So, its effect is not detrimental as long as the propagation direction of each laser is the same for the excitation and deexcitation of a certain Rydberg state. For this reason, we assume $z_0=0$ to analyze the phase modulation because of the atomic drift. The phases of the Rabi frequencies for pumping $|r_1\rangle$ and $|r_2\rangle$ are always opposite to each other, so that during $t\in (0,~\pi/\sqrt2\Omega]$ we have
\begin{eqnarray}
 C_{\text{r1}}(t) &=& -iC_{\text{r}}(t) e^{i\varphi(t)},~C_{\text{r2}}(t)= -iC_{\text{r}}(t) e^{-i\varphi(t)},\label{varphi01}
\end{eqnarray}
where $C_{\text{r}}(t)$ is a positive variable, and $\pi/\sqrt2\Omega$ is the time required for a complete depopulation of the ground state. The factor $-i$ in the above equation arises from the usual optical pumping with a constant Rabi frequency. To see what determines the phase $\varphi(t)$ at $t= \pi/\sqrt2\Omega$, the solid curve in Fig.~\ref{figure003} shows the value of $\varphi(\pi/\sqrt2\Omega)$ as a function of $v$ when $\Omega/2\pi=\sqrt2$~MHz. It shows that $\varphi(t)$ at $t= \pi/\sqrt2\Omega$ is linear with $v$. Indeed, the dashed curve in Fig.~\ref{figure003} shows that $\varphi(\pi/\sqrt2\Omega) /  \frac{2\pi kv}{\Omega}$ is $0.1287$ for all the values of $v$ when $k =k_-$. For $k =k_+$, there is a minor change of $\varphi(\pi/\sqrt2\Omega) /  \frac{2\pi kv}{\Omega}$ from $0.1287$ to $0.1298$ when $v$ increases from $0.005$ to $0.1$m/s; however, this change is negligible, and is much less prominent for larger $\Omega$. For instance, with $\Omega/2\pi =4\sqrt2$~MHz, $\varphi(\pi/\sqrt2\Omega) /  \frac{2\pi kv}{\Omega}$ becomes $0.1287$ for all the values of $v$ with the same parameters in both Fig.~\ref{figure003}(a) and~\ref{figure003}(b). This means that the phase modulation $\varphi$ in the Rydberg states $|r_{1(2)}\rangle$ is almost linear in the atomic speed. We note that a similar linear phase modulation in a transition between a ground and a Rydberg state is used in Ref.~\cite{Shi2019Dopp2} for a similar purpose. 

The reason to ignore the velocity change of the qubit upon absorption of photon in Eq.~(\ref{Hamil01}) is as follows. We take a rubidium-87 atom as an example. The velocity change of an atom by absorbing a photon with $k_\pm $ is $\delta v_\pm \approx 0.015~(0.004)$m/s. During $t\in (0,~\pi/\sqrt2\Omega]$, the total time for the atom to experience the velocity change of  $\delta v_\pm$, i.e, to stay in the Rydberg state $|r_1\rangle$ or $|r_2\rangle$, is only $\pi/(4\sqrt2\Omega)$. The value of $\varphi(\pi/\sqrt2\Omega) $ is $0.002$~rad with $v=0.005$~m/s in Fig.~\ref{figure003}(a), and $0.029$~rad with $v=0.015$~m/s in Fig.~\ref{figure003}(b), respectively. This means that the recoil effect will induce an extra phase change of less than $0.0005~(0.0072)$~rad if $k=k_-~(k_+)$. This effect is already quite minor, and can be further suppressed if an $\Omega$ larger than that employed in Fig.~\ref{figure003} is used. A further effect of the recoil is entanglement of the spatial location and internal state of the atom. However, for $t_{\text{w}}\sim1\mu$s, the separation of the $|r_1\rangle$-part and the $|r_2\rangle$-part of the atom during the wait period is about 15 and $4$~nm with $k_+$ and $k_-$, respectively, which will not cause detrimental effect upon the subsequent optical pumping. However, if a laboratory can only achieve an $\Omega$ that is much smaller than the one used in Fig.~\ref{figure003}, the method of Sec.~\ref{sec02a} should be used that comes with no recoil effect.

The linear form of $\varphi(\pi/\sqrt2\Omega) $ is an important character. It means that the Doppler dephasing is in fact quite regular, which points to a general method to eliminate its detrimental effect as discussed below.

\subsubsection{Optimal duration for state restoration}
We then examine conditions to bring the state back to the ground state soon after the qubit is pumped to the Rydberg states $|r_{1(2)}\rangle$ at $t=\pi/\sqrt2\Omega$. The state restoration is accompanied with the cancellation of the phase $\varphi$ accumulated during $(0,~\pi/\sqrt2\Omega]$. Because the signs of $\varphi$ during the excitation and deexcitation of Rydberg states are opposite to each other, the phase cancellation should be possible. Nevertheless, if we continue to use a Rabi frequency $\Omega_{\text{dp}}$ for Rydberg depopulation that is equal to the Rabi frequency $\Omega$, state restoration within a time of $\pi/\sqrt2\Omega$ is not optimal. This is because with $\Omega_{\text{dp}}=\Omega$, full restoration can only happen if the atom suddenly changes its velocity along $\mathbf{z}$ at the onset of the Rydberg depopulation so that a ``time-reversal'' process of the phase accumulation happens. Unfortunately, this is not possible for an atom in free flight.

\begin{figure}
\includegraphics[width=3.3in]
{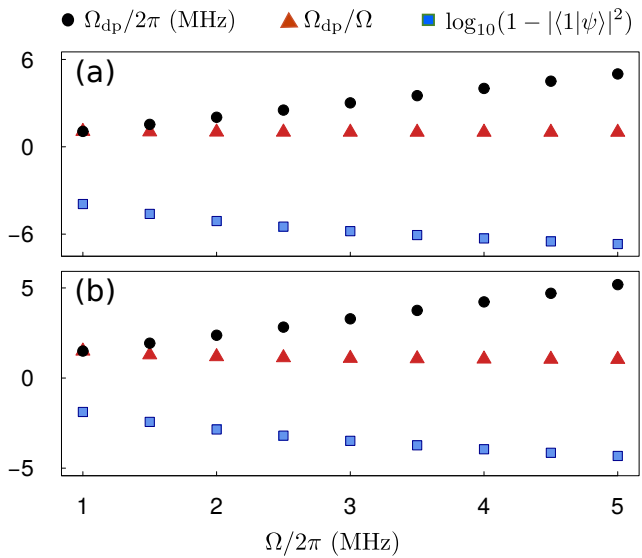}
\caption{ The circular symbols show the optimal values of $\Omega_{\text{dp}}$ for state restoration after populating the Rydberg state. $k=k_-$ and $k_+$ in (a) and (b), respectively. Triangular symbols show the ratio between $\Omega_{\text{dp}}$ and $\Omega$, and the square symbols show the population error~(by common logarithm) in the ground state after using $\Omega_{\text{dp}}$ for one and half Rabi cycles. Here $v = 0.05$m/s. \label{figure04} }
\end{figure}
  To proceed, we note that the phase factor in the Rabi frequency during Rydberg depopulation has an extra term of $\pm \pi k v/\sqrt2\Omega$ compared with those of Eq.~(\ref{Hamil01}). Because $| \pi  k v/\sqrt2\Omega|> |\varphi(\pi/\sqrt2\Omega)|$, this extra phase not only compensates the phase $\varphi(\pi/\sqrt2\Omega) $ accumulated during the Rydberg excitation, but also induces an extra phase $\varphi_{\text{e}}= \pi k v\sqrt2\Omega-\varphi(\pi/\sqrt2\Omega)$. This extra $\varphi_{\text{e}}$ should be compensated during the Rydberg deexcitation~(in the opposite direction). We numerically find that the Rydberg deexcitation by one and half Rabi cycles results in a better state restoration, similar to the case studied in Fig.~\ref{figure02}(e); also, we find that the optimal Rabi frequency $\Omega_{\text{dp}}$ for deexcitation is slightly different from $\Omega$. Because both the phase accumulations and the extra phase $\pm k v\sqrt2\Omega$ in the Rabi frequencies are linear in the atomic velocity, an optimal duration for bringing the population back to ground state with one value of $v$ is also optimal for other values of $v$. Then, we choose $v= 0.05$~m/s to numerically locate the best duration, $3\pi/\sqrt2\Omega_{\text{dp}}$, for the Rydberg depopulation, which is determined by an optimal Rabi frequency $\Omega_{\text{dp}}$. The numerically found values of $\Omega_{\text{dp}}$, its ratio to $\Omega$, and the population error in the ground state are shown in Fig.~\ref{figure04}. For $(\Omega,~\Omega_{\text{dp}})/2\pi=(2,~2.0288)$~MHz as an example, Fig.~\ref{figure04}(a) shows that the final population error in $|1\rangle$ is $7.9\times10^{-6}$ when $k=k_-$. For $k=k_+$, the error is a little larger, changing from $1.3\times10^{-2}$ to $4.8\times10^{-5}$ for $\Omega/2\pi\in[1,~5]$~MHz, shown in Fig.~\ref{figure04}(b).

For the implementation of a Rydberg blockade gate, it is necessary to induce a $\pi$ phase shift to the ground state when the population is restored. Then, one can use a negative $\Omega_{\text{dp}}$ for the deexcitation of the Rydberg state. The optimal value of $|\Omega_{\text{dp}}|$ for a given $\Omega$ is different from the value of $\Omega_{\text{dp}}$ shown in Fig.~\ref{figure04}. For example, $\Omega_{\text{dp}}/2\pi$ is respectively $1.0577$ and $1.5394$ when $\Omega/2\pi = 1$ and $1.5$~MHz in Fig.~\ref{figure04}(a). But for $\Omega_{\text{dp}}<0$, we find that the optimal value of $\Omega_{\text{dp}}/2\pi$ becomes $-1.0674$ and $-1.5460$~MHz for these two cases, respectively.

\begin{figure}
\includegraphics[width=3.2in]
{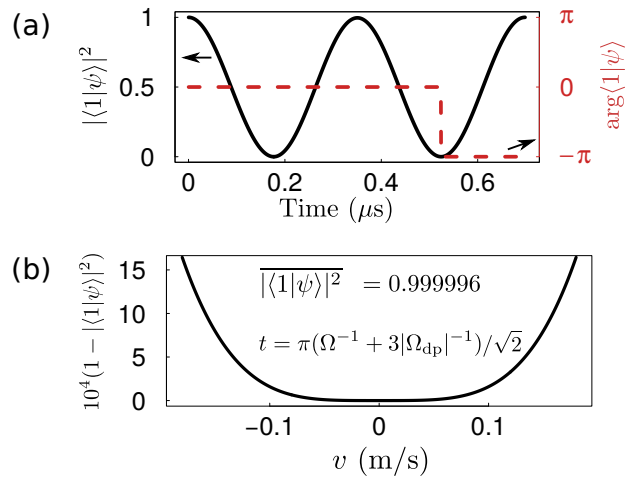}
\caption{(a) Time evolution of the population~(solid curve) and phase~(dashed curve) of the component $|1\rangle$ during the quantum control by the Hamiltonian of Eq.~(\ref{Hamil01}), where $\Omega/2\pi$ is $2$~MHz during $t\in[0,~\pi/(\sqrt2\Omega))$, and $\Omega_{\text{dp}}/2\pi=-2.0399$~MHz during $t\in[0,~3\pi/|\sqrt2\Omega_{\text{dp}}|]+\pi/(\sqrt2\Omega)$. The atomic velocity is $v=0.05$~m/s. The final population error is $1-|\langle1|\psi\rangle|^2=1.0\times10^{-5}$. (b) Population error in the ground state with the same Hamiltonian but for different $v$. When $T=10~\mu$K, the average population error is $4\times10^{-6}$. In both (a) and (b), $k=k_-$ is used; the final phase is $\pi$~(or, equivalently, $-\pi$, according to the data) in the ground state, i.e., no phase error appears according to the precision of the computers used for simulation. \label{fig_05} }
\end{figure}

 As an example, Fig.~\ref{fig_05}(a) shows the time evolution of the ground-state component of the wavefunction when the initial state is $|1\rangle$ with $\Omega/2\pi =2$~MHz. With an atomic speed of $v=0.05$~m/s and the optimal $\Omega_{\text{dp}}/2\pi=-2.0399$~MHz for state restoration, the final error $1-|\langle1|\psi\rangle|^2$ at the end of the pulse is $1.0\times10^{-5}$, and no phase error is found. For different $v$, the final population error is shown in Fig.~\ref{fig_05}(b), and no phase error is found, too. When the data in Fig.~\ref{fig_05}(b) is averaged over the Maxwell distribution of the atomic velocity with $T=10~\mu$K, the average population error in $|1\rangle$ is below $10^{-5}$, and the averaged phase is $\pi$ with no error. 
 
\section{Case 2: When there is a gap time}\label{sec03}

A necessary step in the Rydberg blockade gate~\cite{PhysRevLett.85.2208} is to pump the control atom from $|1\rangle$ to Rydberg states, let it remain there for a time $t_{\text{w}}$, and then pump it back to the ground state. As studied in Refs.~\cite{Wilk2010,Saffman2011,Saffman2016,Shi2019Dopp2}, the atomic motion during the gap time $t_{\text{w}}$ contributes a major Doppler dephasing to the quantum process. As shown in Sec.~\ref{sec02}, the different atomic velocity imprints different phase shifts to the Rydberg states. This amounts to a {\it random} energy shift to the atom, and thus seems impossible to be eliminated by any linear method after the gap time. Below, we present a theory that is able to suppress this randomly distributed energy shift.

\subsection{Why should there be a wait time of $t_{\text{w}}$? }
Before presenting the detail of the theory, it is useful to clarify the role played by the wait time. In the three-pulse protocol of a controlled-phase gate by using Rydberg blockade~\cite{PhysRevLett.85.2208}, the key step is to let one of the two qubits, the control qubit, stay in a Rydberg state if it is initialized in an appointed qubit state~(supposing it is $|1\rangle$), during which the other qubit, the target atom, is pumped by a resonant light pulse to a Rydberg state. Because of the Rydberg blockade, the target atom can't go to the Rydberg state if the control qubit is already in the Rydberg state, leading to an input-state-dependent phase shift to the two-qubit system. This protocol does not require fine tuning of the magnitude of the blockade energy as long as the blockade is much larger than the Rabi frequency of the resonant pumping. Then, if we let the state of the control qubit oscillate between different nearby Rydberg states, the Rydberg blockade between the control and target qubits remains. In this sense, as long as the control atom~(when initialized in $|1\rangle$) is in any of the Rydberg states during the Rydberg excitation of the target qubit, the Rydberg blockade gate can be implemented.

\subsection{State restoration after the wait}
Based on the above observation, we study the state restoration when there is a gap between excitation and deexcitation of the Rydberg states. As soon as the population is pumped to the Rydberg states $|r_{1(2)}\rangle$ at $t=\pi/\sqrt2\Omega$, a time $t_{\text{w}}$ is allowed to elapse, during which the atom stays at the Rydberg states. Meanwhile, the atom continues to drift. Then, the atomic location along $\mathbf{z}$ is $z_0+v(\pi/\sqrt2\Omega+ t_{\text{w}})$ at the beginning of the optical pumping for bringing its state back to the ground state. This means that if we continue to use the Rabi frequency $\Omega$ for optical pumping, it can not restore the population back to the ground state as in Fig.~\ref{figure02}. It is because that state restoration requires the cancellation of the phase $\varphi$ accumulated during $(0,~\pi/\sqrt2\Omega]$, but unfortunately the extra phase $\pm kvt_{\text{w}}$ in the Rabi frequency will induce an extra phase accumulation as an error. For a waiting period $t_{\text{w}}$ near 1 microsecond, the phase $\pm kvt_{\text{w}}$ is much larger than the $\varphi$ shown in Fig.~\ref{figure003}(a), which means that whatever time is used for the state restoration, it is impossible to compensate the phase $\varphi$. Moreover, the wait time $t_{\text{w}}$ is the same for all possible drift speeds, resembling a linear process put in a random precession that will surely result in a severe dephasing. These two factors lead to a large damping to the ground-Rydberg coherence because of the wait time~\cite{Saffman2011}. 

\begin{table*}[ht]
  \centering
  \begin{tabular}{|c|c|c|c|c|c|}
    \hline   \text{Method}&     $(\Omega,~\Omega_{\text{dp}})/2\pi$~(MHz)&  $t_{\text{w}}$~($\mu$s)  & T~($\mu$K)& $\overline{|\langle1|\psi\rangle|^2} $ &  $\overline{|\text{arg}\langle1|\psi\rangle |}$  \\ \hline
\text{This work}&	 $(2.0,~-2.0339)$  &   \multirow{2}{*} {\large{$\frac{\sqrt2}{2}$}} &  \multirow{2}{*}{10}  &  0.9999797  & $ \pi$	 \\  \cline{1-2}\cline{5-6}
\text{Traditional method}&$(2\sqrt2,~2\sqrt2)$ & 	   &  & 0.9999955 & 3.024902\\\hline
\text{This work}&	 $(2.0,~-2.0339)$  &    \multirow{2}{*} {\large{$\frac{\sqrt2}{2}$}}  &  \multirow{2}{*}{200}  &  0.9968510  & $ \pi$	 \\  \cline{1-2}\cline{5-6} 
\text{Traditional method}&$(2\sqrt2,~2\sqrt2)$ &	   & & 0.9984545 & 2.620949\\\hline
\text{This work}&	 $(2.0,~-2.0339)$  &  \multirow{2}{*} {$\sqrt2$}  &  \multirow{2}{*}{200} &  0.9922810  & $ \pi$	 \\ \cline{1-2}\cline{5-6}  
\text{Traditional method}&$(2\sqrt2,~2\sqrt2)$ & 	   &  & 0.9961266 & 2.208995
\\ \hline 
  \end{tabular}
  \caption{  \label{table1} Performance of state restoration with the same wait time for our method with the setup in Fig.~\ref{figure06} and the traditional method. In our method, infrared laser fields are used to induce the transition $|r_{1(2)}\rangle\leftrightarrow|r_3\rangle$ during the wait time as shown in Fig.~\ref{figure06}(b), where $|r_3\rangle$ plays a role similar to that played by $|1\rangle$ supposing $|1\rangle$ is pumped. $\overline{(\cdots)}$ denotes the average of the population or phase of the component $|1\rangle$ in the wavefunction at the end of the pulse sequence, which is calculated by using Maxwell distribution at a temperature of $T$. No error in the phase is found for our method, while the error of the phase by the traditional method is significant, shown in the last column of the table. $z_0=0$ is used here; deviation of $z_0$ from $0$ has no effect on the traditional method, but will decrease the population error in our method~(see Fig.~\ref{figure09}). }
  \end{table*}


\begin{figure}
\includegraphics[width=3.3in]
{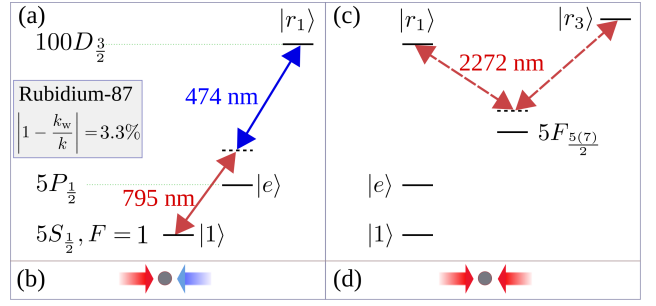}
\caption{(a) Laser excitation of the $d$-orbital Rydberg state $|r_1\rangle$ from the qubit state $|1\rangle=|5S_{1/2},F=1,m_F=1\rangle$ via an intermediate state in the $5P_{1/2}$ manifold of $^{87}$Rb; the other qubit state is $|5S_{1/2},F=2,m_F=2\rangle$. The lower and upper circular polarized laser lights propagate oppositely to each other along the quantization axis, so that the effective wavevector is $k=k_- = 2\pi(1/474-1/795)$nm$^{-1}$. The $D_{5/2}$ Rydberg state is not excited because of the selection rules. (b) During the wait time, the Rydberg states $|r_{1(2)}\rangle$ are pumped to another Rydberg state $|r_3\rangle$ via the intermediate state $5F_{5/2}$; both fine states $5F_{5/2}$ and $5F_{7/2}$ are shown because of the energy separation between them is only $0.6$~GHz~\cite{Sansonetti2006}. The two infrared laser lights counterpropagate with each other, so that the effective wavevector is $k_{\text{w}}=(4\pi/2272)$nm$^{-1}$.  \label{figure06} }
\end{figure}

\begin{figure}
\includegraphics[width=2.5in]
{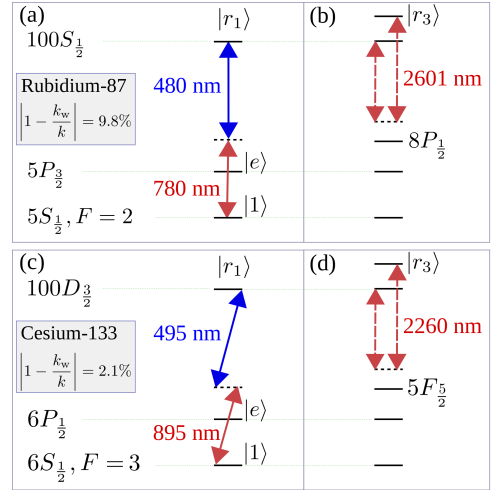}
\caption{(a) Excitation of $s$-orbital Rydberg states of $^{87}$Rb by using $5P_{3/2}$ as the intermediate state. $|1\rangle=|5S_{1/2},F=2,m_F=2\rangle$, and the fields are $\pi$ polarized along the quantization axis. (b) $|r_{1(2)}\rangle$ are pumped to $|r_3\rangle$ via the intermediate state $8P_{1/2}$ during the wait time; the fine splitting between $8P_{1/2}$ and $8P_{3/2}$ is $0.57$~THz~\cite{Sansonetti2006}. The effective wavevector is $k_{\text{w}}=(4\pi/2601)$nm$^{-1}$. (c) and (d) are similar to (a) and (b), but are the case when $|6P_{1/2},F=4\rangle$ serves as the intermediate state for exciting $d$-orbital Rydberg states of $^{133}$Cs. Here $|1\rangle=|6S_{1/2},F=3,m_F=3\rangle$ and the fields are right-hand polarized. The fine splitting between $5F_{7/2}$ and $5F_{5/2}$ is $4.5$~GHz~\cite{Sansonetti2009}.  \label{figure0602} }
\end{figure}

\begin{figure}
\includegraphics[width=2.5in]
{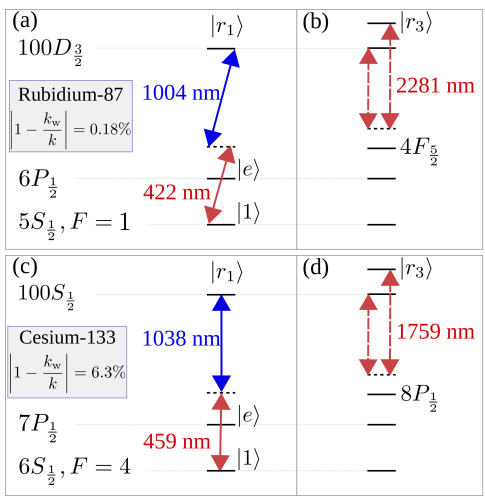}
\caption{Scheme by using higher low-lying intermediate states for the excitation of $|r_{1(2)}\rangle$  in our theory. (a) Excitation of $d$-orbital Rydberg states of $^{87}$Rb by using $6P_{1/2}$ as $|e\rangle$. Here, $|1\rangle$ is $|5S_{1/2},F=1,m_F=1\rangle$ and the fields are circularly polarized. The wavelengths for the transitions in (a) are $421.7$ and $1003.6$~nm, while that in (b) is $2281.4$~nm. (c) and (d) show contents similar to those in (a) and (b), where $7P_{1/2}$ and $8P_{1/2}$ of $^{133}$Cs are used as intermediate states in (c) and (d), respectively, where all fields are $\pi$ polarized. \label{figure0603} }
\end{figure}

We tackle the above issue by using another Rydberg state, $|r_3\rangle$, to serve as a pseudo qubit state during the gap time; in other words, it plays the role of the qubit state $|1\rangle$ although it is a Rydberg state near $|r_1\rangle$ and $|r_2\rangle$. The criteria to choose an appropriate $|r_3\rangle$ is that the wavevector of the fields for the transitions $|r_{1(2)}\rangle\leftrightarrow |1\rangle$ should be nearly equal to that for $|r_{1(2)}\rangle \leftrightarrow |r_3\rangle$. When $k=k_-$ is used for the Rydberg excitation stage, the configuration in Fig.~\ref{figure06} can fulfill this condition, where $|r_{1(2)}\rangle$ is a $D_{3/2}$ state of a large principal quantum number. For instance, $|r_{1(2)}\rangle$ can be $|100(101)D_{3/2}, m_J = 3/2, m_I=3/2\rangle$. During the Rydberg excitation, the lower~(upper) light field has a wavelength of $795.0~(473.9)$~nm~\cite{Sansonetti2006}, so that the wavevector is $k=k_-\approx5.35\times10^{6}/m$. Specifically, the wavevectors for $|1\rangle\rightarrow|r_1\rangle$ is $k_-$, while that for $|1\rangle\rightarrow|r_2\rangle$ is $-k_-$. During the wait period, two infrared laser fields are used to connect $|r_1\rangle$ and $|r_3\rangle$, where $|r_3\rangle$ can be either a $D_{3/2}$ Rydberg state, or a $g$-orbital Rydberg state; similarly, two counterpropagating infrared fields are used to connect $|r_2\rangle$ and $|r_3\rangle$. The configuration is chosen so that the wavevector for $|r_3\rangle\rightarrow|r_1\rangle$ is $k_{\text{w}}$, while that for $|r_3\rangle\rightarrow|r_2\rangle$ is $-k_{\text{w}}$. When the two infrared light fields propagate oppositely to each other along the quantization axis, the wavevector is then $k_{\text{w}}=4\pi/2271.8$~nm$^{-1}\approx5.53\times10^{6}/m$ for the setup in Fig.~\ref{figure06}(b)~\cite{Sansonetti2006}. The difference between the effective wavevector, $k_-$, during the Rydberg excitation and that during the wait period, $k_{\text{w}}$, is about $3.3\%$, which is small enough for removing the Doppler dephasing as shown below.

The configuration in Fig.~\ref{figure06} is not the only choice when rubidium-87 is used in our method. For instance, if the intermediate state $5P_{3/2}$ is used in the Rydberg excitation, the configuration in Fig.~\ref{figure0602}(a) and~\ref{figure0602}(b) can be used. There are several differences between Fig.~\ref{figure06} and Fig.~\ref{figure0602}(a,b): in the former~(latter) case, the qubit state $|1\rangle$ is $|5S_{1/2},~F=1(2),~m_F=1(2)\rangle$, the laser fields are circularly~(linearly) polarized, the Rydberg state is a $d~(s)-$orbital state, and the mismatch between $k_-$ and $k_{\text{w}}$ is $3.3~(9.8)\%$. If cesium-133 is used, we find that the configuration in Fig.~\ref{figure0602}(c) and~\ref{figure0602}(d) is useful, where $|1\rangle=|6S_{1/2},~F=3,~m_F=3\rangle$ and right-hand polarized light fields are used to induce the transition $|1\rangle\rightarrow|nD_{3/2},~m_J=3/2,~m_I=7/2\rangle$, where $n$ is a large principal quantum number. The values of $k$ and $k_{\text{w}}$ are respectively $2\pi(1/494.6-1/894.6)$ and $4\pi/2260.5$~nm$^{-1}$~\cite{Sansonetti2009} in Fig.~\ref{figure0602}(c,d), with a very small mismatch of $2.1\%$, which is the best case if the lowest $p$-orbital state is used as the intermediate state. The next best case for cesium-133 is to choose $6P_{3/2}$ as $|e\rangle$ and $9P_{1/2}$ as the intermediate state for the infrared laser fields, with $|1-k_{\text{w}}/k|\approx5.2\%$. 

If higher low-lying intermediate states are used, we find that the configurations in Fig.~\ref{figure0603} are good cases. For the case of rubidium-87, one can choose $6P_{1/2}$ for the excitation of Rydberg states. Then, if $4F_{5/2}$ is used during the gap time for the pumping between $|r_3\rangle$ and $|r_{1(2)}\rangle$, the wavevectors in Fig.~\ref{figure0603}(a) and in~\ref{figure0603}(a) have a mismatch of $0.18\%$. In fact, if lower Rydberg states with principal quantum numbers of $76$ or $77$ are used, $|1-k_{\text{w}}/k|$ is smaller than $10^{-4}$. If $6P_{1/2}$ is replaced by $6P_{3/2}$ in Fig.~\ref{figure0603}(a), $|1-k_{\text{w}}/k|$ becomes $0.94\%$, which is still quite small. For $^{133}$Cs, the best case is shown in Figs.~\ref{figure0603}(c,d) if $7P_{1/2}$ state is used for the excitation of Rydberg state, where $|1-k_{\text{w}}/k|$ is $6.3\%$, which becomes $7.6\%$ if $8P_{3/2}$ is used in Fig.~\ref{figure0603}(d), or $9.0\%$ if $7P_{3/2}$ is used in Fig.~\ref{figure0603}(c). 

Comparing the values of $|1-k_{\text{w}}/k|$ in different configurations, the configuration in Fig.~\ref{figure0602}(c,d) is best for $^{133}$Cs, and the one in Fig.~\ref{figure0603}(a,b) is the best for $^{87}$Rb, although the latter seems better. But because the mass of a $^{133}$Cs atom is about 1.5 times that of the $^{87}$Rb, the cesium atoms can have smaller drifting velocity. To show that our method can be used even if the mismatch between $k_-$ and $k_{\text{w}}$ is not smallest and the qubit is not as heavy as a cesium qubit, we choose, for example, the configuration in Fig.~\ref{figure06} for a numerical study. 


\begin{figure}
\includegraphics[width=3.2in]
{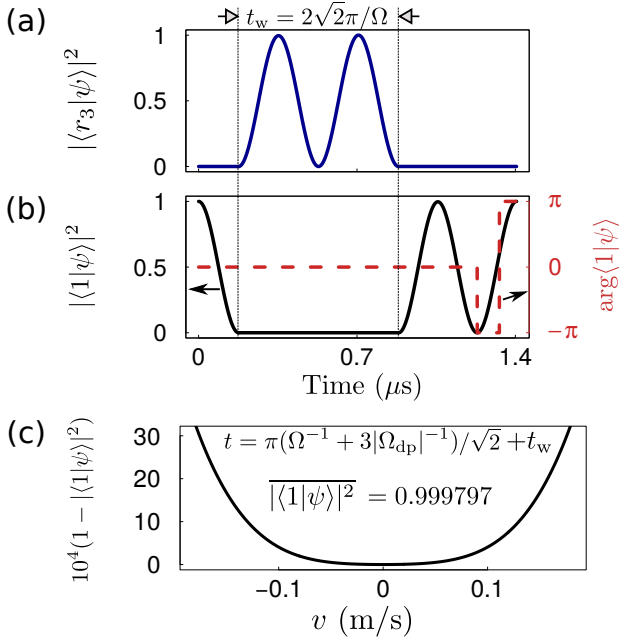}
 \caption{State restoration with a wait time of $t_{\text{w}}=2\sqrt2\pi/\Omega$ inserted between the excitation and deexcitation of Rydberg states with the setup in Fig.~\ref{figure06}. Similar to Fig.~\ref{fig_05}, the Rabi frequencies are $\Omega/2\pi=2~$MHz for the Rydberg excitation, and $\Omega_{\text{dp}}/2\pi=-2.0339~$MHz for the Rydberg deexcitation. (a) Population evolution in $|r_3\rangle$ during the time $t\in[\frac{\pi}{\sqrt2 \Omega},~\frac{5\pi}{\sqrt2 \Omega})$ when the infrared laser fields induce state transfer between $|r_3\rangle$ and $|r_{1(2)}\rangle$, as shown in Fig.~\ref{figure06}(b). (b) Time evolution of the population and phase of the ground-state component $|1\rangle$ in the wavefunction. The value of $v=0.05$~m/s is used in both (a) and (b). (c) Population error in $|1\rangle$ at the end of the pumping as a function of $v$. When $T=10~\mu$K, the averaged population error in $|1\rangle$ is $2.0\times10^{-4}$. $z_0=0$ is used in (a)-(c). \label{figure07} }
\end{figure}

\begin{figure}
\includegraphics[width=3.3in]
{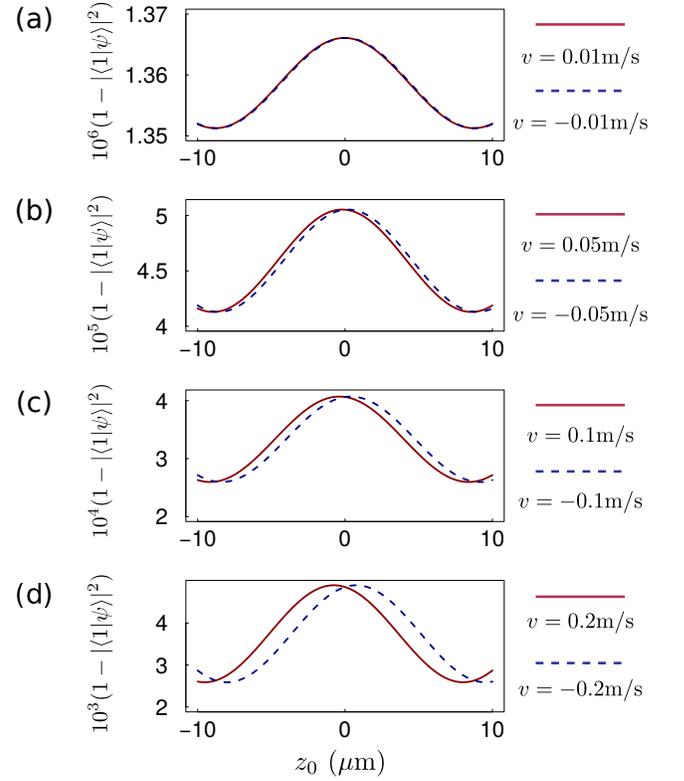}
 \caption{Dependence of the final population error on the initial coordinate, $z_0$, of the atom along $\mathbf{z}$ with the setup in Fig.~\ref{figure06} when infrared laser fields are used in the gap time. The atomic speeds along $\mathbf{z}$ are $0.01,~0.05,~0.1,~0.2$~m/s in (a), (b), (c), and (d), respectively. The results are scaled up by $10^{n}$, with $n=6,~5,~4$, and $3$ in (a), (b), (c), and (d), respectively; the phase of the final state is $\pi$ with no error. Solid~(dashed) curves show results with positive~(negative) speeds; the solid and dashed curves are symmetrical to each other with respect to the line of $z_0=0$. Laser parameters and the wait duration are the same as those in Fig.~\ref{figure07}(a) and~\ref{figure07}(b).  \label{figure09} }
\end{figure}

Figures~\ref{figure07}(a) and~\ref{figure07}(b) show results for the state restoration with a wait time of $t_{\text{w}}=4\pi/(\sqrt2\Omega) =\sqrt2/2~\mu$s, atomic velocity $v=0.05$~m/s, and initial qubit location $z_0=0$ along $\mathbf{z}$ of a $^{87}$Rb atom. During the wait time, infrared laser fields are used to induce a transition between $|r_3\rangle$ and $|r_{1(2)}\rangle$ via a largely detuned intermediate state $5F_{5/2}$~($5F_{7/2}$ is not coupled because of the selection rules). The field for the transition $|r_3\rangle\leftrightarrow|5F_{5/2}\rangle$ propagates oppositely to that for the transition $|r_{1(2)}\rangle\leftrightarrow|5F_{5(7)/2}\rangle$, so that the wavevector of the Rabi frequency is $\pm k_{\text{w}}$. After the Rydberg excitation at $t= \pi/(\sqrt2\Omega)$, the optical laser fields cease, and infrared laser fields are switched on, with a Rabi frequency equal to that used in the Rydberg excitation. The population oscillates between $|r_3\rangle$ and $|r_{1(2)}\rangle$, as shown in Fig.~\ref{figure07}(a); the population branching between $|r_{1}\rangle$ and $|r_{2}\rangle$ is not shown because they depend on the value of $z_0$, which is chosen as $0$ in Fig.~\ref{figure07}. Although there is a $3.3\%$ difference between $k_-$ and $k_{\text{w}}$, the result in Fig.~\ref{figure07}(a) shows that the population in $|r_3\rangle$ is $9.2\times10^{-6}$ at the end of the gap time, demonstrating that $|r_3\rangle$ almost plays the same role of $|1\rangle$ supposing $|1\rangle$ is instead pumped during the gap time. After the gap time, infrared lasers are switched off, and Rydberg deexcitation lasers are switched on, with $\Omega_{\text{dp}}/2\pi=-2.0339~$MHz, as used in Fig.~\ref{fig_05}. After one and half Rabi cycles, the population returns to the qubit state $|1\rangle$ with an error of $5.0\times10^{-5}$, where arg$\langle1|\psi\rangle = \pi$, i.e., no phase error appears. We also simulate the time evolution of the wavefunction for different atomic velocities when $z_0=0$, where Fig.~\ref{figure07}(c) shows the final scaled population error, $10^4(1-|\langle1|\psi\rangle |^2)$, for each of them. Using the Maxwell distribution with $T=10~\mu$K, the average value of $|\langle1|\psi\rangle |^2$ is about $0.9998$.

A distinct feature of the setup in Figs.~\ref{figure06}(c) and~\ref{figure06}(d) is that the final population error in the ground state $|1\rangle$ differs for different $z_0$. This phenomenon is quite different from the case where no gap time is allowed: the results in Figs.~\ref{figure02}-\ref{fig_05} do not change when $z_0$ changes~(except of an extra constant phase shift in Fig.~\ref{figure003}). To put it in perspective, Fig.~\ref{figure09} shows the final population error for different $z_0$ when the qubit velocity along $\mathbf{z}$ applies the values of $0.01,~0.05,~0.1,~0.2$~m/s, as shown in Figs.~\ref{figure09}(a), \ref{figure09}(b), \ref{figure09}(c), and \ref{figure09}(d), respectively. The data are obtained by using the same parameters~(except of $z_0$ and $v$) as used in Figs.~\ref{figure07}(a) and~\ref{figure07}(b). The reason for different error to appear when $z_0$ differs lies in that there is a $3.3\%$ difference between the wavevector of the optical laser fields, $k_-$, and that of the infrared laser fields, $k_{\text{w}}$. This difference violates the condition that $|r_3\rangle$ serves as a perfect pseudo-qubit state. A striking feature in Fig.~\ref{figure09} is that for $|v|\lesssim0.1$~m/s, the larger $|z_0|$ is, the less significant will the final error be for $|z_0|<10~\mu$m. The position fluctuation of the qubit is less than $10~\mu$m in typical experimental setups~\cite{Isenhower2010}, and so we can assume $z_0=0$ to obtain an upper bound of the error. Here, it is necessary to clarify the meaning of $z_0=0$: it refers to that when the qubit drifts from the center of the optical trap, the Rabi frequencies at the moment of $t$ for the transition $|1\rangle\rightarrow|r_1\rangle$ is $\Omega e^{ikvt}$ and that for $|1\rangle\rightarrow|r_2\rangle$ is $\Omega e^{-ikvt}$. In other words, it requires the factor $ e^{i\zeta+i\mathbf{k}\cdot \mathbf{r}}$ of the Rabi frequencies from the forward-propagating fields and that from the backward-propagating fields to be equal, where $\mathbf{r}$ denotes the coordinate of the trap center, and $\zeta$ is a phase factor in the atomic dipole moment determined by the selection rules. This requirement can be satisfied either by adjustment of the phase of the laser oscillators or by using linear optics.  

Our theory can also apply to hotter qubits and longer wait times. For instance, if the qubits are only cooled to $200~\mu$K, the characteristic atomic velocity of the qubit becomes $v=\sqrt{k_BT/m}\approx0.14$m/s. Then, the average of the error should be evaluated by taking a larger range of $v$ than those shown in Fig.~\ref{figure07}(c); the convergence is verified by that the integration of the Maxwell distribution should be 1. Then, we find that the average value of $|\langle1|\psi\rangle |^2$ is $0.9969$ with a wait time of $t_{\text{w}}=\sqrt2/2~\mu$s and atomic temperature $200~\mu$K. If even longer wait time is used, the population is $0.9923$ with a wait time of $t_{\text{w}}=8\pi/(\sqrt2\Omega) =\sqrt2~\mu$s at $200~\mu$K, where four Rabi cycles between $|r_3\rangle$ and $|r_{1(2)}\rangle$ are used during the gap time. In all these results, the phase of $\langle1|\psi\rangle$ at the end of the sequence is $\pi$ with no error observed. 

A comparison between our theory and the traditional method is shown in Table~\ref{table1}, where the traditional pumping refers to the method in Fig.~\ref{figure01}(b). In order to show an unbiased comparison, we use a Rabi frequency of $\Omega/2\pi = 2\sqrt2$~MHz in the traditional method so that its time for a $\pi$ pulse is the same as that in our method. The data for the traditional method were obtained as follows: use a $\pi$ pulse with Rabi frequency $\Omega e^{ik(z_0+vt)}$ for the transition $|1\rangle\leftrightarrow|r_1\rangle$, wait for a time of $t_{\text{w}}$, and use another $\pi$ pulse with the Rabi frequency $\Omega e^{ik(z_0+vt)}$ for the transition $|1\rangle\leftrightarrow|r_1\rangle$. Table~\ref{table1} shows that although the population errors in $|1\rangle$ are of similar magnitude in our theory and the traditional method, the phase error in the latter method is significant. In contrast, our theory has completely suppressed the phase error within the precision of our computer processor~(the calculated phase error is below $10^{-10}$). We note that the phase error can lead to large error for the entanglement generation. For example, Ref.~\cite{Saffman2011} estimated that the entangling gate has an error of about $0.15$ for a gap time of $t_{\text{w}}=1.4~\mu$s if $T=200~\mu$K~(see Fig.~3 of Ref.~\cite{Saffman2011}) which used a wavevector of $2\pi(1/480-1/780)$nm$^{-1}$ that is similar to the one used here.

\subsection{Application in Rydberg blockade gate}\label{sec03C}
We show that the method above can be used to suppress the Doppler dephasing error in an entangling gate with neutral Rydberg atoms, which is a major stumbling block for achieving an accurate two-qubit Rydberg gate~\cite{Saffman2011}. Ideally, the Rydberg blockade gate maps the input states according to $\{|00\rangle,~|01\rangle,~|10\rangle,~|11\rangle\}\rightarrow \{|00\rangle,~-|01\rangle,~-|10\rangle,~-|11\rangle\}$~\cite{PhysRevLett.85.2208}. Due to the finiteness of the blockade interaction and Rydberg-state decay, the actual gate in the matrix form becomes,
\begin{eqnarray}
 \mathscr{U} &=& \left(
  \begin{array}{cccc}
    1& 0 & 0&0\\
    0 & a &0&0\\
    0 &0 & b&0\\
    0& 0 & 0&c\\   
    \end{array} 
  \right) ,\nonumber
  \end{eqnarray}
where $a$ and $b$ can deviate from $-1$ due to Doppler dephasing, $c$ can obtain errors from both Doppler dephasing and the finiteness of the blockade interaction, and the magnitudes of $a,~b$, and $c$ become smaller than 1 because of Rydberg-state decay. Then, the fidelity error is given by
\begin{eqnarray}
 E &=& E_{\text{ro}} + E_{\text{decay}}.\nonumber
\end{eqnarray}
Here the rotation error is~\cite{Pedersen2007}
\begin{eqnarray}
 E_{\text{ro}} &=& 1-\frac{1}{20}\left[  |\text{Tr}(U^\dag \mathscr{U})|^2 + \text{Tr}(U^\dag \mathscr{U}\mathscr{U}^\dag U ) \right], \label{fidelityError01}
\end{eqnarray}
where $\mathscr{U}$ is evaluated by using the Hamiltonian dynamics with the Rydberg-state decay ignored, $U=$ diag$\{1,-1,-1,-1\}$, and the Rydberg-state decay can be approximated by~\cite{Zhang2012} 
\begin{eqnarray}
E_{\text{decay}}&=&[T_{\text{r}}(01)+T_{\text{r}}(10) +T_{\text{r}}(11)]/(4\tau),\label{fidelityError02}
\end{eqnarray}
where $T_{\text{r}}(ab)$ is the time for the input state $|ab\rangle$ to be in a single-Rydberg state; the time for the input state $|11\rangle$ to be in a two-atom Rydberg state $|r_\alpha r_\beta\rangle$ throughout the gate sequence is tiny and neglected. We let $\tau$ be the smallest among all the lifetimes of the Rydberg states, which will slightly overestimate the Rydberg-state decay.

A numerical estimate of Eq.~(\ref{fidelityError01}) can be proceeded by integrating the Schr\"odinger equation for each input state. In the setup of Fig.~\ref{figure06}, the $d$-orbital Rydberg state is excited from $|5S_{1/2},F=1,m_F=1\rangle$. We choose, as an example, the following Rydberg states: $|r_1\rangle= |95D_{3/2},~m_J= 3/2,~m_I=3/2\rangle$,~$|r_2\rangle= |97D_{3/2},~m_J= 3/2,~m_I=3/2\rangle$,~$|r_3\rangle= |99D_{3/2},~m_J= 3/2,~m_I=3/2\rangle$, and consider a geometry of setup used in the experiment of Ref.~\cite{Isenhower2010}, where the quantization axis is along $\mathbf{z}$, and the two centers of the traps used for the two qubits are at $(0,~0,~0)$ and $(L,~0,~0)$, which is a useful configuration for the blockade gate. By using the quantum defects reported in Refs.~\cite{Li2003,Han2006}, the van der Waals coefficient, $C_6$, of the diagonal energy shifts for the states $|r_\alpha r_\beta\rangle$ can be calculated~\cite{Walker2008} to be $[-14,~-21,~29,~-18,~-26]~$THz~$\mu m^6$ when the principal quantum numbers of the state $(|r_\alpha\rangle,~|r_\beta\rangle)$ are $[(95,~95),~(95,~97),~(95,~99),~(97,~97)~(97,~99)]$, respectively. A two-atom Rydberg state $|r_\alpha r_\beta\rangle$ is also coupled to other two-atom Rydberg states because the total angular momentum is not conserved. These processes can be neglected because the chance to excite both atoms to Rydberg state is negligible. The energy separation between $|r_1\rangle$ and $|r_{2(3)}\rangle$ is $16~(30)$~GHz, which is large enough to distinguish them with different laser fields. The decay error is estimated with the lifetime of $|r_1\rangle$ which is $\tau=787~\mu$s in a temperature of $4~\mu$K~\cite{Beterov2009}.

\begin{table*}[ht]
  \centering
  \begin{tabular}{|c|c|c|c|c|c|}
    \hline   \text{Method}&     $\frac{1}{2\pi}(\Omega,~\Omega_{\text{dp}},~\Omega_{\text{t}},~\Omega_{\text{IF}})$~(MHz)& T~($\mu$K)&  $t_{\text{w}}$~($\mu$s)&  \text{Gate duration}~($\mu$s)  &  $\overline{E_{\text{ro}}} $~[Eq.~(\ref{roterror02})]  \\ \hline
\text{This work}&	 $(2.0,~-2.0339,~2,~2)$& \multirow{2}{*}{10}  & \multirow{2}{*} {\large{$\frac{\sqrt2}{2}$}} &1.405     & $2.56\times{10^{-4}}$  	 \\   \cline{1-2}\cline{5-6}
\text{Traditional gate}&$(2\sqrt2,~2\sqrt2,~2\sqrt2,~-)$ & 	   & & 1.061 & $4.69\times{10^{-3}}$   \\\hline
\text{This work}&	 $(2.0,~-2.0339,~2,~2)$& \multirow{2}{*}{200}  & \multirow{2}{*} {\large{$\frac{\sqrt2}{2}$}} &1.405   &  $1.99\times{10^{-3}}$ 	 \\ \cline{1-2}\cline{5-6}
\text{Traditional gate}&$(2\sqrt2,~2\sqrt2,~2\sqrt2,~-)$ & 	   & & 1.061 & $8.06\times{10^{-2}}$ \\\hline
\text{This work}&	 $(2.0,~-2.0339,~2,~2)$& \multirow{2}{*}{10}  & \multirow{2}{*} {$\sqrt2$} &2.111   & $6.64\times{10^{-4}}$  	 \\  \cline{1-2}\cline{5-6}
\text{Traditional gate}&$(2\sqrt2,~2\sqrt2,~2\sqrt2,~-)$ & 	   & & 1.768 & $1.41\times{10^{-2}}$ \\\hline
\text{This work}&	 $(2.0,~-2.0339,~2,~2)$& \multirow{2}{*}{200}  & \multirow{2}{*} {$\sqrt2$} &2.111   & $5.58\times{10^{-3}}$ 	 \\  \cline{1-2}\cline{5-6}
\text{Traditional gate}&$(2\sqrt2,~2\sqrt2,~2\sqrt2,~-)$ & 	   & & 1.768 & $2.03\times{10^{-1}}$ \\
\hline
  \end{tabular}
  \caption{  \label{table2} Performance of the blockade gate when $t_{\text{w}}=\sqrt2/2~(\sqrt2)~\mu$s for both our method and the traditional method. The setup in Fig.~\ref{figure06} is used in our method, where the Rabi frequencies of the infrared lasers and for Rydberg excitation and deexcitation of the control qubit are $\Omega_{\text{IF}},~\Omega$, and $\Omega_{\text{dp}}$, respectively; the Rabi frequency for the target qubit is $\Omega_{\text{t}}$ and $-\Omega_{\text{t}}$ for the first $\pi$ pulse and the latter $3\pi$ pulse~(for the induction of a $\pi$ phase, see Fig.~\ref{figure07}). The wait time is given by $t_{\text{w}}=2\sqrt2\pi/\Omega_{\text{IF}}$~(or $4\sqrt2\pi/\Omega_{\text{IF}})$. The traditional method is implemented with the well known $\pi-2\pi(\text{gap})-\pi$ pulse sequence with all Rabi frequencies equal to $\Omega$. $|1\rangle$ is excited to $|r_{1(2)}\rangle$~($|r_{1}\rangle$) in our~(the traditional) method; see text above Eq.~(\ref{hamiltonian0}) for details. Rydberg interactions are calculated with a relatively small $L=7~\mu$m so as to demonstrate that our theory can suppress the Doppler dephasing which dominates the rotation error. We assume $z_{0,c}=z_{0,t}=0$ for the initial location of the two qubits; variation of $z_0$ can decrease the rotation error for our method~(see Fig.~\ref{figure09}). }
  \end{table*}


During the gate sequence, the control qubit, if initialized in $|1\rangle$, is pumped to $|r_{1(2)}\rangle$ with a $\pi$ pulse. Then, there is a wait time $t_{\text{w}}$, during which the target qubit, if initialized in $|1\rangle$, experiences ``1+3'' $\pi$ pulses, shown in Fig.~\ref{fig_05}(a), and meanwhile the control qubit experiences $n$ times $4\pi$ pulses of the infrared laser fields as shown in Figs.~\ref{figure06}(c) and~\ref{figure06}(d)~[Fig.~\ref{figure07}(a) shows the case when $n=1$]. The duration of the pulses for the target qubit should be shorter than or equal to the wait time; we choose $t_{\text{w}}=4n\pi/(\sqrt2\Omega)$, where $n=1$ or $2$ here. After the wait time, three $\pi$ pulses are used to deexcite the control qubit, as shown in Figs.~\ref{figure06}(a),~\ref{figure06}(b), and~\ref{figure07}(b). Rotation error for the input state $|01\rangle$ is evaluated by using the method as in Fig.~\ref{fig_05}, and that for $|10\rangle$ is evaluated by using the calculation used in Fig.~\ref{figure07}. The input state $|11\rangle$ is analyzed as following. At the beginning of the wait time, the population is distributed in $|11\rangle,~|r_11\rangle,~|r_21\rangle$. The $|11\rangle$ part evolves in a way similar to $|01\rangle$ during the wait time, while the $|r_11\rangle$ and $|r_21\rangle$ components of the wavefunction evolve under the following Hamiltonian,
\begin{widetext}
 \begin{eqnarray}
  \centering
\frac{1}{2}  \left(
  \begin{array}{ccccccccc}
    2V_{23}& 0& \Omega_{\text{t}}e^{-ikz_t} & \Omega_{\text{\tiny{IF}}} e^{ik_{\text{w}}z_c}& 0& 0 &  \Omega_{\text{\tiny{IF}}} e^{-ik_{\text{w}}z_c} &0&0\\
    0& 2V_{13}& \Omega_{\text{t}}e^{ikz_t}& 0 & \Omega_{\text{\tiny{IF}}} e^{ik_{\text{w}}z_c}& 0 & 0 &  \Omega_{\text{\tiny{IF}}} e^{-ik_{\text{w}}z_c} &0\\
     \Omega_{\text{t}} e^{ikz_t}&\Omega_{\text{t}} e^{-ikz_t}&0& 0& 0  & \Omega_{\text{\tiny{IF}}} e^{ik_{\text{w}}z_c}& 0 & 0 &  \Omega_{\text{\tiny{IF}}} e^{-ik_{\text{w}}z_c} \\
   \Omega_{\text{\tiny{IF}}} e^{-ik_{\text{w}}z_c}  &0& 0&2V_{22}& 0& \Omega_{\text{t}}e^{-ikz_t}  &  0&0&0\\
   0&  \Omega_{\text{\tiny{IF}}} e^{-ik_{\text{w}}z_c} & 0& 0&2V_{12}& \Omega_{\text{t}}e^{ikz_t}  &  0&0&0\\
  0&  0&  \Omega_{\text{\tiny{IF}}} e^{-ik_{\text{w}}z_c} & \Omega_{\text{t}} e^{ikz_t}&\Omega_{\text{t}} e^{-ikz_t}&0 &  0&0&0\\
   \Omega_{\text{\tiny{IF}}} e^{ik_{\text{w}}z_c}  &0& 0& 0& 0  &  0&2V_{12}&0& \Omega_{\text{t}}e^{-ikz_t}\\
   0&  \Omega_{\text{\tiny{IF}}} e^{ik_{\text{w}}z_c} & 0& 0& 0 &  0&0&2V_{11}& \Omega_{\text{t}}e^{ikz_t} \\
  0&  0&  \Omega_{\text{\tiny{IF}}} e^{ik_{\text{w}}z_c} & 0& 0&0 &  \Omega_{\text{t}} e^{ikz_t}&\Omega_{\text{t}} e^{-ikz_t}&0
    
    \end{array}
  \right),
  \label{hamiltonian0}
 \end{eqnarray}
\end{widetext}
 where $z_c=z_{0,c}+v_ct$ and $z_t=z_{0,t}+v_tt$ are the coordinates along $\mathbf{z}$ of the control and target qubits, respectively, $z_{0,c(t)}$ is the coordinate at the beginning of the gate sequence, and $v_{c(t)}$ is the velocity of the control~(target) qubit along $\mathbf{z}$. The interactions are given by $[V_{11},~V_{12},~V_{13},~V_{22},~V_{23}]= [C_6(r_1r_1),~C_6(r_1r_2),~C_6(r_1r_3),~C_6(r_2r_2),~C_6(r_2r_3)]/L^6$, and the basis of the above matrix is $|r_3r_2\rangle,~|r_3r_1\rangle$,$~|r_31\rangle$,~$|r_2r_2\rangle,~|r_2r_1\rangle,~|r_21\rangle$,~$|r_1r_2\rangle,~|r_1r_1\rangle$,$~|r_11\rangle$. $\Omega_{\text{t}}$ is the Rabi frequency for the transition $|1\rangle\leftrightarrow|r_{1(2)}\rangle$ of the target qubit, and $\Omega_{\text{\tiny{IF}}}$ is the Rabi frequency for the transition $|r_3\rangle\leftrightarrow|r_{1(2)}\rangle$ of the control qubit. $\Omega_{\text{\tiny{IF}}} $ lasts for $4n\pi$ pulses, but $\Omega_{\text{t}}$ lasts for ``1+3'' $\pi$ pulses, shown in Fig.~\ref{fig_05}, where its signs in the first $\pi$ and the latter $3\pi$ pules are opposite. In the numerical simulation, we use $z_{0,c}=z_{0,t}=0$ so as to obtain an upper bound for the error because the state-transfer error is smallest with this condition, shown in Fig.~\ref{figure09}. Also, we have ignored the change of the Rydberg interactions because of the atomic drift during the gate sequence. This is mainly because the blockade mechanism is robust against the small fluctuation of the Rydberg interaction.

For comparison, we also study the fidelity error for a traditional gate, which is simulated as follows: (i) use a $\pi$ pulse in the control qubit for the transition $|1\rangle\rightarrow|r_1\rangle$, (ii) use a $2\pi$ pulse in the target qubit for the transition $|1\rangle\rightarrow|r_1\rangle\rightarrow|1\rangle$, (iii) wait for a time $t_{\text{w}}-2\pi/\Omega'$, and (iv) use a $\pi$ pulse in the control qubit for the transition $|r_1\rangle|\rightarrow1\rangle$, where all Rabi frequencies are $\Omega'=\sqrt2\Omega$. Here, we note that in practice, the wait time $t_{\text{w}}-2\pi/\Omega'$ is usually divided into two equal pieces that are symmetrically placed before and after the $2\pi$ pulses, as in Ref.~\cite{Graham2019}; the method used in step (iii) is for the sake of simplicity. With this process, the time of the gap between the excitation and deexcitation of the control qubit is equal to the one used in the gate by our theory. 
 
The evaluation of the average rotation error, $\overline{E_{\text{ro}}}$, involves a two-dimensional integration over the distribution of $v_c$ and $v_t$, both of which apply the Maxwell distribution $\mathscr{G}(v)$~(which is Gaussian in one dimension). Because an accurate convergence of this integration requires long time of simulation, we approximate the two-dimensional integration by
\begin{eqnarray}
\overline{E_{\text{ro}}}\approx \frac{\sum_{v_c}\sum_{v_t} E_{\text{ro}}(v_c,~v_t)\mathscr{G}(v_c) \mathscr{G}(v_c) }{ \sum_{v_c}\sum_{v_t} \mathscr{G}(v_c) \mathscr{G}(v_c)},\label{roterror02}
\end{eqnarray}
where the sum is over $10^4$ sets of speeds $(v_{c},~v_t)$, where $v_{c(t)}$ applies $100$ values equally distributed from $-0.5$ to $0.5$~m/s because the atomic speed has little chance to be over $0.5$~m/s for $T<0.2$~mK. According to numerical calculation, our method suppresses the rotation error by about 20~(40) times compared with the traditional method when $T=10~(200)~\mu$K whether $t_{\text{w}}$ is $\sqrt2/2$ or $\sqrt2~\mu$s, shown in Table~\ref{table2}. This shows that our method is specifically useful for hotter qubits. The values of $\overline{E_{\text{ro}}}$ for the traditional gate in Table~\ref{table1} have similar magnitude with the estimate shown in Fig.~3 of Ref.~\cite{Saffman2011}.

The results in Table~\ref{table2} show that it is possible to achieve a high-fidelity blockade gate by our method when $t_{\text{w}}=\sqrt2/2~\mu$s. The error of our gate from Rydberg-state decay is determined by Eq.~(\ref{fidelityError02}), which is $E_{\text{decay}}\approx 7\sqrt2\pi/(4\Omega\tau)$ by using the data in Table~\ref{table2} where $|\Omega_{\text{dp}}|\approx\Omega$. This leads to $E_{\text{decay}}\approx7.86\times10^{-4}$ with the lifetime of the Rydberg state given by that of the $95d$ state of rubidium, $\tau\approx787~\mu$s. Thus, the fidelity can reach $\mathcal{F}=1-\overline{E_{\text{ro}}}-E_{\text{decay}}\approx0.999~(0.997)$ when $T=10~(200)~\mu$K with our method. With the parameters in Table~\ref{table2}, the fidelity is $\mathcal{F}\approx0.995~(0.919)$ when $T=10~(200)~\mu$K with the traditional method. This means that our method can offer a blockade gate with fidelity~$0.999$ with atoms cooled only to $10~\mu$K. If qubits are cooled beyond $10~\mu$K, motional dephasing can be further suppressed so that the rotation error becomes the fundamental blockade error $E_{\text{ro}}\approx (\sqrt2\Omega/V_{11})^2/8$~\cite{Saffman2005}, which is about $10^{-4}$ with the parameters used here. 

Table~\ref{table2} shows that when the wait time changes from $t_{\text{w}}=\sqrt2/2$ to $\sqrt2~\mu$s, there is little increase of the rotation error in our method. On the other hand, the rotation error increases significantly for the traditional method. However, because the large wait time, the decay error in our gate, $E_{\text{decay}}\approx1.24\times10^{-3}$, dominates the fidelity error. So, even with our method, it is still necessary to shorten the wait time so as to shrink the Rydberg-state decay if gate fidelity beyond $0.999$ is desired.

\section{Discussions}\label{sec04}
\subsection{Application in gate protocols with high intrinsic fidelity}\label{sec05A}
The study of the blockade gate in Sec.~\ref{sec03C} shows that the gate fidelity is limited by the Rydberg-state decay and the blockade error even if the motional dephasing is suppressed by our method. To obtain a two-qubit entangling gate with fidelity $\lesssim10^{-4}$ that is necessary for fault-tolerant quantum computing based on measurement-free error correction~\cite{Crow2016}, it is useful to employ our theory to suppress the motional dephasing so as to recover the high intrinsic fidelity in those protocols.

First, one can use a spin-echo sequence to suppress the blockade error so that the Rydberg-state decay becomes the final fundamental limit~\cite{Shi2018prapp2}. The spin-echo method, however, requires that each element in the Hamiltonian reverses their sign before the onset of the `echo' stage. This means that the Rydberg interaction of each two-atom Rydberg state not only changes sign, but also changes the magnitude by a common ratio. Although such a condition can be easily satisfied if only Rydberg eigenstates are excited during the gate sequence, it seems difficult to be satisfied if superposition states of $|r_1\rangle$ and $|r_2\rangle$ are used as required in our theory. Nevertheless, one can use two right-hand circularly polarized light fields to excite $|r_1\rangle$, as in Fig.~\ref{figure06}(a), but the lower~(upper) field for exciting $|r_2\rangle$ can be right-hand~(left-hand) circularly polarized. In this case, $|r_1\rangle$ and $|r_2\rangle$ are $D_{3/2}$ Rydberg states of the same principal quantum number and can share the same $m_F$ but have different $m_I$. To make sure either $|r_1\rangle$ or $|r_2\rangle$ is a Rydberg eigenstate, external fields can be used to shift away irrelevant Zeeman sublevels~\cite{Isenhower2010,Levine2019}. Then, an atom in either $|r_1\rangle$ or $|r_2\rangle$ exhibits the same interaction with another Rydberg atom. Furthermore, $|r_{3}\rangle$ can be a superposition state~\cite{ShiJPB2016,Shi2017pra,Shi2018a} stabilized by strong microwave fields, where the interaction coefficient can be continuously tuned by adjustment of the microwave fields as analyzed in Ref.~\cite{Shi2017pra}; this can finally lead to that an atom in $|r_{1}\rangle$ or $|r_{2}\rangle$ or $|r_{3}\rangle$ has the same interaction with another Rydberg atom so that the required condition for spin echo can be satisfied with the scheme in this work. This method can be used in configurations other than that in Fig.~\ref{figure06}, too.

A second route toward a two-qubit entangling gate with high intrinsic fidelity is via quantum interference between different Rabi cycles of Rydberg atoms~\cite{Shi2018Accuv1}. A Rydberg eigenstate needs to be excited in the original gate protocol of Ref.~\cite{Shi2018Accuv1}; this is the same requirement as in Ref.~\cite{Shi2018prapp2}, thus can be solved by the same method as discussed above. Similar scheme is applicable to the gate protocols in~\cite{Shi2017} which has no blockade error.

Third, there are several Rydberg gate protocols based on pulse shaping or stimulated Raman adiabatic passage~\cite{Goerz2014,Theis2016,Petrosyan2017,Yu2019,Shen:19,Liao2019}. The reason for these gates to have high fidelity is that the time-dependent Rabi frequency can suppress the blockade error, but the motional dephasing is still there. Thus, it is possible to use the theory in this work to recover the high intrinsic fidelity for the gate protocols in~\cite{Goerz2014,Theis2016,Petrosyan2017,Yu2019,Shen:19,Liao2019}.

\subsection{Application in multiqubit gates}
The method in this work can also be used in multiqubit quantum gate based on Rydberg blockade. For example, Ref.~\cite{Isenhower2011} proposed a controlled-NOT gate where $k$ atoms serve as the control unit, with $k$ up to $50$. Whether the sequential or the simultaneous version of the gate in Ref.~\cite{Isenhower2011} is implemented, it will suffer from motional dephasing in any gap time. Similarly, the three-qubit Deutsch and Toffoli gates in Ref.~\cite{Shi2018prapp} can also benefit from the theory in this work. The three-qubit gates in~\cite{Li2018,Beterov2018arX,Su2018}, however, require excitation of Rydberg eigenstates, and thus can not use the method in this work directly unless the method in Sec.~\ref{sec05A} is employed.

\subsection{Comparison with the protocol in Ref.~\cite{Shi2019Dopp2}}
Besides this work~(denoted as Paper I), Ref.~\cite{Shi2019Dopp2}~(denoted as Paper II) has recently presented another protocol for a similar purpose. In addition to their distinct physics, there are several differences between Paper I and Paper II.

First, the protocol in Paper I can work either with a gap time or not, while the protocol in Paper II always comes with a gap time when the state of the atom changes back and forth between two Rydberg eigenstates. Regarding their application in quantum gates, the theory in both Paper I and Paper II can be used for the traditional Rydberg blockade gate which has gap times, but only the theory in Paper I can be directly used for high-fidelity gate implemented by a single laser pulse as proposed in Ref.~\cite{Shi2018Accuv1} and discussed in Sec.~\ref{sec05A}. Recently, Ref.~\cite{Levine2019} reported a fast gate protocol that can be understood as an extension of the $U_1$ gate of Ref.~\cite{Shi2018Accuv1}. In~\cite{Levine2019}, a phase shift $\xi$ is inserted into the Rabi frequency at the middle of the gate sequence, which leads to a condition $2\alpha-\beta=\pi$ in the two angles $\alpha$ and $\beta$ of the $U_1$ gate in~\cite{Shi2018Accuv1}. Because there is a finite transient time to induce the phase shift which effectively breaks the pulse into two pieces, optimization as in~\cite{Levine2019} may be necessary if the theory in Paper I is used in such a case. 

Second, the method in Paper I needs four sets of laser fields to excite the Rydberg state for each qubit, while the protocol in Paper II only needs two, as in the conventional two-photon excitation of Rydberg states~\cite{Wilk2010,Isenhower2010,Zhang2010,Maller2015,Zeng2017,Levine2018,Picken2018,Levine2019,Graham2019}. This means that the protocol in Paper II is relatively simpler to implement.

Third, the magnitudes of Rabi frequencies used in different stages of the protocol in Paper II can not be equal and should satisfy a condition determined by the wavelengths of the laser fields, while those in Paper I can be equal. On the other hand, the methods in both papers work well only when $kv/\Omega\ll1$~(a figure of merit of 0.1 should be good enough), where $v$ is a typical velocity of the qubit. As a consequence, if $\Omega$ should be bounded below a certain value, the method in Paper I can have a better performance especially when qubits are cooled only to the submillikelvin regime. For qubits cooled to around $10~\mu$K, the methods in both papers are comparable.

Fourth, the method in Paper I can use the $6p$~($7p$) state~(henceforth referred to as $|e\rangle$) for the excitation of Rydberg states in $^{87}$Rb~($^{133}$Cs), and meanwhile the intermediate state used during the gap time~( referred to as $|e'\rangle$) can be even higher than $|e\rangle$, as shown in Fig.~\ref{figure0603}. On the other hand, at least one of $|e\rangle$ and $|e'\rangle$ should be the lowest $p$ state in the method of Paper II. Because the dipole moment between $|e(e')\rangle$ and a high-lying Rydberg state is small, it is desirable to use higher $|e(e')\rangle$ for achieving larger Rabi frequencies so that shorter gate operation is obtained that comes with less Doppler dephasing. However, the dipole moment between $|1\rangle$ and $|e\rangle$ also decays quickly if $|e\rangle$ goes up. So a compromise is to choose the second $p$-orbital state, i.e., $6p$~($7p$) state as $|e\rangle$ for $^{87}$Rb~($^{133}$Cs), as done in recent experiments of Rydberg gates~\cite{Maller2015,Levine2018,Levine2019,Graham2019}. This means that compared with the method in Paper II that is restricted to smaller Rabi frequencies, the method in Paper I can be faster if configurations as in Fig.~\ref{figure0603} are employed.

According to the differences described above, it should be useful to implement the method in Paper II in the control qubit for the Rydberg blockade gate. Although the target qubit still suffers from dephasing, it does not have any gap time so that one can get a fidelity larger than that of the traditional gate. For further improvement, the method in Paper I can be used in the target qubit to suppress the dephasing in it. Of course, it should be easier to use the theory in either Paper I or Paper II for all qubits in a large-scale quantum processor.

\section{Conclusions}\label{sec05}
We introduce a theory to suppress the motion-induced dephasing of the transition between ground and Rydberg states of a flying neutral atom. The theory relies on using two counterpropagating sets of fields to excite two Rydberg states, $|r_1\rangle$ and $|r_2\rangle$. The Rabi frequencies for $|r_1\rangle$ and $|r_2\rangle$ are $\Omega e^{ik(z_0+vt)}$ and $(\Omega e^{-ik(z_0+vt)})$, respectively, where the atomic speed along the light propagation is $v$. This method results in a high-fidelity excitation of Rydberg state by one $\pi$ pulse, and an accurate deexcitation of Rydberg state by a $3\pi$ pulse. When there is a gap time between the excitation and deexcitation of the Rydberg states, infrared laser fields can be used to induce a transition between $|r_{1(2)}\rangle$ and another Rydberg state $|r_3\rangle$ that is near $|r_{1(2)}\rangle$. Because $|r_1\rangle,~|r_{2}\rangle$, and $|r_3\rangle$ are all Rydberg states, the atom can still block the Rydberg excitation of a nearby atom during the gap time, so that the presence of the infrared laser fields has no detrimental effect on the blockade mechanism. We find that one can restore the state from Rydberg to ground states after the gap time when the wavevectors for the Rydberg excitation~(deexcitation) and that of the infrared laser fields are equal, which is a condition that can be satisfied with a negligible error for both rubidium and cesium. Numerical simulation shows that a Rydberg blockade gate by this method can suppress the dephasing error by more than one order of magnitude compared to the gate by the traditional pumping method with the same gap time. Then, it becomes possible to recover the intrinsic fidelity of the blockade gate with atoms cooled only to the order of $10~\mu$K. This theory breaks the fundamental coherence limit set by the atomic motion  in a number of applications by using Rydberg atoms. It also brings opportunity to use quantum interference to suppress decoherence in a general quantum system.

\section*{ACKNOWLEDGMENTS}
The author thanks Yan Lu for insightful discussions. This work was supported by the National Natural Science Foundation of China under Grant No. 11805146.


%



\begin{thebibliography}{58}%
\makeatletter
\providecommand \@ifxundefined [1]{%
 \@ifx{#1\undefined}
}%
\providecommand \@ifnum [1]{%
 \ifnum #1\expandafter \@firstoftwo
 \else \expandafter \@secondoftwo
 \fi
}%
\providecommand \@ifx [1]{%
 \ifx #1\expandafter \@firstoftwo
 \else \expandafter \@secondoftwo
 \fi
}%
\providecommand \natexlab [1]{#1}%
\providecommand \enquote  [1]{``#1''}%
\providecommand \bibnamefont  [1]{#1}%
\providecommand \bibfnamefont [1]{#1}%
\providecommand \citenamefont [1]{#1}%
\providecommand \href@noop [0]{\@secondoftwo}%
\providecommand \href [0]{\begingroup \@sanitize@url \@href}%
\providecommand \@href[1]{\@@startlink{#1}\@@href}%
\providecommand \@@href[1]{\endgroup#1\@@endlink}%
\providecommand \@sanitize@url [0]{\catcode `\\12\catcode `\$12\catcode
  `\&12\catcode `\#12\catcode `\^12\catcode `\_12\catcode `\%12\relax}%
\providecommand \@@startlink[1]{}%
\providecommand \@@endlink[0]{}%
\providecommand \url  [0]{\begingroup\@sanitize@url \@url }%
\providecommand \@url [1]{\endgroup\@href {#1}{\urlprefix }}%
\providecommand \urlprefix  [0]{URL }%
\providecommand \Eprint [0]{\href }%
\providecommand \doibase [0]{http://dx.doi.org/}%
\providecommand \selectlanguage [0]{\@gobble}%
\providecommand \bibinfo  [0]{\@secondoftwo}%
\providecommand \bibfield  [0]{\@secondoftwo}%
\providecommand \translation [1]{[#1]}%
\providecommand \BibitemOpen [0]{}%
\providecommand \bibitemStop [0]{}%
\providecommand \bibitemNoStop [0]{.\EOS\space}%
\providecommand \EOS [0]{\spacefactor3000\relax}%
\providecommand \BibitemShut  [1]{\csname bibitem#1\endcsname}%
\let\auto@bib@innerbib\@empty
\bibitem [{\citenamefont {Saffman}\ \emph {et~al.}(2010)\citenamefont
  {Saffman}, \citenamefont {Walker},\ and\ \citenamefont
  {{M{\o}lmer}}}]{Saffman2010}%
  \BibitemOpen
  \bibfield  {author} {\bibinfo {author} {\bibfnamefont {M.}~\bibnamefont
  {Saffman}}, \bibinfo {author} {\bibfnamefont {T.~G.}\ \bibnamefont {Walker}},
  \ and\ \bibinfo {author} {\bibfnamefont {K.}~\bibnamefont {{M{\o}lmer}}},\
  }\bibfield  {title} {\bibinfo {title} {{Quantum information with Rydberg
  atoms}},\ }\href@noop {} {\bibfield  {journal} {\bibinfo  {journal} {Rev.
  Mod. Phys.}\ }\textbf {\bibinfo {volume} {82}},\ \bibinfo {pages} {2313}
  (\bibinfo {year} {2010})}\BibitemShut {NoStop}%
\bibitem [{\citenamefont {Saffman}(2016)}]{Saffman2016}%
  \BibitemOpen
  \bibfield  {author} {\bibinfo {author} {\bibfnamefont {M.}~\bibnamefont
  {Saffman}},\ }\bibfield  {title} {\bibinfo {title} {{Quantum computing with
  atomic qubits and Rydberg interactions: Progress and challenges}},\
  }\href@noop {} {\bibfield  {journal} {\bibinfo  {journal} {J. Phys. B}\
  }\textbf {\bibinfo {volume} {49}},\ \bibinfo {pages} {202001} (\bibinfo
  {year} {2016})}\BibitemShut {NoStop}%
\bibitem [{\citenamefont {Weiss}\ and\ \citenamefont
  {Saffman}(2017)}]{Weiss2017}%
  \BibitemOpen
  \bibfield  {author} {\bibinfo {author} {\bibfnamefont {D.~S.}\ \bibnamefont
  {Weiss}}\ and\ \bibinfo {author} {\bibfnamefont {M.}~\bibnamefont
  {Saffman}},\ }\bibfield  {title} {\bibinfo {title} {{Quantum computing with
  neutral atoms}},\ }\href {\doibase 10.1063/PT.3.3626} {\bibfield  {journal}
  {\bibinfo  {journal} {Phys. Today}\ }\textbf {\bibinfo {volume} {70}},\
  \bibinfo {pages} {44} (\bibinfo {year} {2017})}\BibitemShut {NoStop}%
\bibitem [{\citenamefont {Dudin}\ and\ \citenamefont
  {Kuzmich}(2012)}]{Dudin2012}%
  \BibitemOpen
  \bibfield  {author} {\bibinfo {author} {\bibfnamefont {Y.~O.}\ \bibnamefont
  {Dudin}}\ and\ \bibinfo {author} {\bibfnamefont {A.}~\bibnamefont
  {Kuzmich}},\ }\bibfield  {title} {\bibinfo {title} {{Strongly interacting
  Rydberg excitations of a cold atomic gas.}}\ }\href@noop {} {\bibfield
  {journal} {\bibinfo  {journal} {Science}\ }\textbf {\bibinfo {volume}
  {336}},\ \bibinfo {pages} {887} (\bibinfo {year} {2012})}\BibitemShut
  {NoStop}%
\bibitem [{\citenamefont {Peyronel}\ \emph {et~al.}(2012)\citenamefont
  {Peyronel}, \citenamefont {Firstenberg}, \citenamefont {Liang}, \citenamefont
  {Hofferberth}, \citenamefont {Gorshkov}, \citenamefont {Pohl}, \citenamefont
  {Lukin},\ and\ \citenamefont {Vuleti{\'c}}}]{Peyronel2012}%
  \BibitemOpen
  \bibfield  {author} {\bibinfo {author} {\bibfnamefont {T.}~\bibnamefont
  {Peyronel}}, \bibinfo {author} {\bibfnamefont {O.}~\bibnamefont
  {Firstenberg}}, \bibinfo {author} {\bibfnamefont {Q.-Y.}\ \bibnamefont
  {Liang}}, \bibinfo {author} {\bibfnamefont {S.}~\bibnamefont {Hofferberth}},
  \bibinfo {author} {\bibfnamefont {A.~V.}\ \bibnamefont {Gorshkov}}, \bibinfo
  {author} {\bibfnamefont {T.}~\bibnamefont {Pohl}}, \bibinfo {author}
  {\bibfnamefont {M.~D.}\ \bibnamefont {Lukin}}, \ and\ \bibinfo {author}
  {\bibfnamefont {V.}~\bibnamefont {Vuleti{\'c}}},\ }\bibfield  {title}
  {\bibinfo {title} {{Quantum nonlinear optics with single photons enabled by
  strongly interacting atoms.}}\ }\href@noop {} {\bibfield  {journal} {\bibinfo
   {journal} {Nature}\ }\textbf {\bibinfo {volume} {488}},\ \bibinfo {pages}
  {57} (\bibinfo {year} {2012})}\BibitemShut {NoStop}%
\bibitem [{\citenamefont {Firstenberg}\ \emph {et~al.}(2013)\citenamefont
  {Firstenberg}, \citenamefont {Peyronel}, \citenamefont {Liang}, \citenamefont
  {Gorshkov}, \citenamefont {Lukin},\ and\ \citenamefont
  {Vuleti{\'c}}}]{Firstenberg2013}%
  \BibitemOpen
  \bibfield  {author} {\bibinfo {author} {\bibfnamefont {O.}~\bibnamefont
  {Firstenberg}}, \bibinfo {author} {\bibfnamefont {T.}~\bibnamefont
  {Peyronel}}, \bibinfo {author} {\bibfnamefont {Q.-Y.}\ \bibnamefont {Liang}},
  \bibinfo {author} {\bibfnamefont {A.~V.}\ \bibnamefont {Gorshkov}}, \bibinfo
  {author} {\bibfnamefont {M.~D.}\ \bibnamefont {Lukin}}, \ and\ \bibinfo
  {author} {\bibfnamefont {V.}~\bibnamefont {Vuleti{\'c}}},\ }\bibfield
  {title} {\bibinfo {title} {{Attractive photons in a quantum nonlinear
  medium}},\ }\href@noop {} {\bibfield  {journal} {\bibinfo  {journal}
  {Nature}\ }\textbf {\bibinfo {volume} {488}},\ \bibinfo {pages} {57}
  (\bibinfo {year} {2013})}\BibitemShut {NoStop}%
\bibitem [{\citenamefont {Li}\ \emph {et~al.}(2013)\citenamefont {Li},
  \citenamefont {Dudin},\ and\ \citenamefont {Kuzmich}}]{Li2013}%
  \BibitemOpen
  \bibfield  {author} {\bibinfo {author} {\bibfnamefont {L.}~\bibnamefont
  {Li}}, \bibinfo {author} {\bibfnamefont {Y.~O.}\ \bibnamefont {Dudin}}, \
  and\ \bibinfo {author} {\bibfnamefont {A.}~\bibnamefont {Kuzmich}},\
  }\bibfield  {title} {\bibinfo {title} {{Entanglement between light and an
  optical atomic excitation.}}\ }\href {\doibase 10.1038/nature12227}
  {\bibfield  {journal} {\bibinfo  {journal} {Nature}\ }\textbf {\bibinfo
  {volume} {498}},\ \bibinfo {pages} {466} (\bibinfo {year}
  {2013})}\BibitemShut {NoStop}%
\bibitem [{\citenamefont {Gorniaczyk}\ \emph {et~al.}(2014)\citenamefont
  {Gorniaczyk}, \citenamefont {Tresp}, \citenamefont {Schmidt}, \citenamefont
  {Fedder},\ and\ \citenamefont {Hofferberth}}]{Gorniaczyk2014}%
  \BibitemOpen
  \bibfield  {author} {\bibinfo {author} {\bibfnamefont {H.}~\bibnamefont
  {Gorniaczyk}}, \bibinfo {author} {\bibfnamefont {C.}~\bibnamefont {Tresp}},
  \bibinfo {author} {\bibfnamefont {J.}~\bibnamefont {Schmidt}}, \bibinfo
  {author} {\bibfnamefont {H.}~\bibnamefont {Fedder}}, \ and\ \bibinfo {author}
  {\bibfnamefont {S.}~\bibnamefont {Hofferberth}},\ }\bibfield  {title}
  {\bibinfo {title} {{Single-Photon Transistor Mediated by Interstate Rydberg
  Interactions}},\ }\href@noop {} {\bibfield  {journal} {\bibinfo  {journal}
  {Phys. Rev. Lett.}\ }\textbf {\bibinfo {volume} {113}},\ \bibinfo {pages}
  {053601} (\bibinfo {year} {2014})}\BibitemShut {NoStop}%
\bibitem [{\citenamefont {Tiarks}\ \emph {et~al.}(2014)\citenamefont {Tiarks},
  \citenamefont {Baur}, \citenamefont {Schneider}, \citenamefont {D{\"u}rr},\
  and\ \citenamefont {Rempe}}]{Tiarks2014}%
  \BibitemOpen
  \bibfield  {author} {\bibinfo {author} {\bibfnamefont {D.}~\bibnamefont
  {Tiarks}}, \bibinfo {author} {\bibfnamefont {S.}~\bibnamefont {Baur}},
  \bibinfo {author} {\bibfnamefont {K.}~\bibnamefont {Schneider}}, \bibinfo
  {author} {\bibfnamefont {S.}~\bibnamefont {D{\"u}rr}}, \ and\ \bibinfo
  {author} {\bibfnamefont {G.}~\bibnamefont {Rempe}},\ }\bibfield  {title}
  {\bibinfo {title} {{Single-Photon Transistor Using a F{\"o}rster
  Resonance}},\ }\href@noop {} {\bibfield  {journal} {\bibinfo  {journal}
  {Phys. Rev. Lett.}\ }\textbf {\bibinfo {volume} {113}},\ \bibinfo {pages}
  {053602} (\bibinfo {year} {2014})}\BibitemShut {NoStop}%
\bibitem [{\citenamefont {Busche}\ \emph {et~al.}(2017)\citenamefont {Busche},
  \citenamefont {Huillery}, \citenamefont {Ball}, \citenamefont {Ilieva},
  \citenamefont {Jones},\ and\ \citenamefont {Adams}}]{Busche2017}%
  \BibitemOpen
  \bibfield  {author} {\bibinfo {author} {\bibfnamefont {H.}~\bibnamefont
  {Busche}}, \bibinfo {author} {\bibfnamefont {P.}~\bibnamefont {Huillery}},
  \bibinfo {author} {\bibfnamefont {S.~W.}\ \bibnamefont {Ball}}, \bibinfo
  {author} {\bibfnamefont {T.}~\bibnamefont {Ilieva}}, \bibinfo {author}
  {\bibfnamefont {M.~P.~A.}\ \bibnamefont {Jones}}, \ and\ \bibinfo {author}
  {\bibfnamefont {C.~S.}\ \bibnamefont {Adams}},\ }\bibfield  {title} {\bibinfo
  {title} {{Contactless nonlinear optics mediated by long-range Rydberg
  interactions}},\ }\href {\doibase 10.1038/nphys4058} {\bibfield  {journal}
  {\bibinfo  {journal} {Nat. Phys.}\ }\textbf {\bibinfo {volume} {13}},\
  \bibinfo {pages} {655} (\bibinfo {year} {2017})}\BibitemShut {NoStop}%
\bibitem [{\citenamefont {Lampen}\ \emph {et~al.}(2018)\citenamefont {Lampen},
  \citenamefont {Nguyen}, \citenamefont {Li}, \citenamefont {Berman},\ and\
  \citenamefont {Kuzmich}}]{Lampen2018}%
  \BibitemOpen
  \bibfield  {author} {\bibinfo {author} {\bibfnamefont {J.}~\bibnamefont
  {Lampen}}, \bibinfo {author} {\bibfnamefont {H.}~\bibnamefont {Nguyen}},
  \bibinfo {author} {\bibfnamefont {L.}~\bibnamefont {Li}}, \bibinfo {author}
  {\bibfnamefont {P.~R.}\ \bibnamefont {Berman}}, \ and\ \bibinfo {author}
  {\bibfnamefont {A.}~\bibnamefont {Kuzmich}},\ }\bibfield  {title} {\bibinfo
  {title} {{Long-lived coherence between ground and Rydberg levels in a
  magic-wavelength lattice}},\ }\href {\doibase 10.1103/PhysRevA.98.033411}
  {\bibfield  {journal} {\bibinfo  {journal} {Phys. Rev. A}\ }\textbf {\bibinfo
  {volume} {98}},\ \bibinfo {pages} {033411} (\bibinfo {year}
  {2018})}\BibitemShut {NoStop}%
\bibitem [{\citenamefont {Adams}\ \emph {et~al.}()\citenamefont {Adams},
  \citenamefont {Pritchard},\ and\ \citenamefont {Shaffer}}]{Adams2019}%
  \BibitemOpen
  \bibfield  {author} {\bibinfo {author} {\bibfnamefont {C.~S.}\ \bibnamefont
  {Adams}}, \bibinfo {author} {\bibfnamefont {J.~D.}\ \bibnamefont
  {Pritchard}}, \ and\ \bibinfo {author} {\bibfnamefont {J.~P.}\ \bibnamefont
  {Shaffer}},\ }\bibfield  {title} {\bibinfo {title} {{Rydberg atom quantum
  technologies}},\ }\href@noop {} {\bibinfo  {journal} {arXiv: 1907.09231}\
  }\BibitemShut {NoStop}%
\bibitem [{\citenamefont {Weimer}\ \emph {et~al.}(2009)\citenamefont {Weimer},
  \citenamefont {M{\"u}ller}, \citenamefont {Lesanovsky}, \citenamefont
  {Zoller},\ and\ \citenamefont {B{\"u}chler}}]{Weimer2009}%
  \BibitemOpen
\bibfield  {journal} {  }\bibfield  {author} {\bibinfo {author} {\bibfnamefont
  {H.}~\bibnamefont {Weimer}}, \bibinfo {author} {\bibfnamefont
  {M.}~\bibnamefont {M{\"u}ller}}, \bibinfo {author} {\bibfnamefont
  {I.}~\bibnamefont {Lesanovsky}}, \bibinfo {author} {\bibfnamefont
  {P.}~\bibnamefont {Zoller}}, \ and\ \bibinfo {author} {\bibfnamefont {H.~P.}\
  \bibnamefont {B{\"u}chler}},\ }\bibfield  {title} {\bibinfo {title} {{A
  Rydberg quantum simulator}},\ }\href {\doibase 10.1038/nphys1614} {\bibfield
  {journal} {\bibinfo  {journal} {Nat. Phys.}\ }\textbf {\bibinfo {volume}
  {6}},\ \bibinfo {pages} {382} (\bibinfo {year} {2009})}\BibitemShut {NoStop}%
\bibitem [{\citenamefont {Jaksch}\ \emph {et~al.}(2000)\citenamefont {Jaksch},
  \citenamefont {Cirac}, \citenamefont {Zoller}, \citenamefont {Rolston},
  \citenamefont {C{\^o}t{\'e}},\ and\ \citenamefont
  {Lukin}}]{PhysRevLett.85.2208}%
  \BibitemOpen
  \bibfield  {author} {\bibinfo {author} {\bibfnamefont {D.}~\bibnamefont
  {Jaksch}}, \bibinfo {author} {\bibfnamefont {J.~I.}\ \bibnamefont {Cirac}},
  \bibinfo {author} {\bibfnamefont {P.}~\bibnamefont {Zoller}}, \bibinfo
  {author} {\bibfnamefont {S.~L.}\ \bibnamefont {Rolston}}, \bibinfo {author}
  {\bibfnamefont {R.}~\bibnamefont {C{\^o}t{\'e}}}, \ and\ \bibinfo {author}
  {\bibfnamefont {M.~D.}\ \bibnamefont {Lukin}},\ }\bibfield  {title} {\bibinfo
  {title} {{Fast Quantum Gates for Neutral Atoms}},\ }\href@noop {} {\bibfield
  {journal} {\bibinfo  {journal} {Phys. Rev. Lett.}\ }\textbf {\bibinfo
  {volume} {85}},\ \bibinfo {pages} {2208} (\bibinfo {year}
  {2000})}\BibitemShut {NoStop}%
\bibitem [{\citenamefont {Wilk}\ \emph {et~al.}(2010)\citenamefont {Wilk},
  \citenamefont {Ga{\"e}tan}, \citenamefont {Evellin}, \citenamefont {Wolters},
  \citenamefont {Miroshnychenko}, \citenamefont {Grangier},\ and\ \citenamefont
  {Browaeys}}]{Wilk2010}%
  \BibitemOpen
  \bibfield  {author} {\bibinfo {author} {\bibfnamefont {T.}~\bibnamefont
  {Wilk}}, \bibinfo {author} {\bibfnamefont {A.}~\bibnamefont {Ga{\"e}tan}},
  \bibinfo {author} {\bibfnamefont {C.}~\bibnamefont {Evellin}}, \bibinfo
  {author} {\bibfnamefont {J.}~\bibnamefont {Wolters}}, \bibinfo {author}
  {\bibfnamefont {Y.}~\bibnamefont {Miroshnychenko}}, \bibinfo {author}
  {\bibfnamefont {P.}~\bibnamefont {Grangier}}, \ and\ \bibinfo {author}
  {\bibfnamefont {A.}~\bibnamefont {Browaeys}},\ }\bibfield  {title} {\bibinfo
  {title} {{Entanglement of Two Individual Neutral Atoms Using Rydberg
  Blockade}},\ }\href@noop {} {\bibfield  {journal} {\bibinfo  {journal} {Phys.
  Rev. Lett.}\ }\textbf {\bibinfo {volume} {104}},\ \bibinfo {pages} {010502}
  (\bibinfo {year} {2010})}\BibitemShut {NoStop}%
\bibitem [{\citenamefont {Saffman}\ \emph {et~al.}(2011)\citenamefont
  {Saffman}, \citenamefont {Zhang}, \citenamefont {Gill}, \citenamefont
  {Isenhower},\ and\ \citenamefont {Walker}}]{Saffman2011}%
  \BibitemOpen
  \bibfield  {author} {\bibinfo {author} {\bibfnamefont {M.}~\bibnamefont
  {Saffman}}, \bibinfo {author} {\bibfnamefont {X.~L.}\ \bibnamefont {Zhang}},
  \bibinfo {author} {\bibfnamefont {A.~T.}\ \bibnamefont {Gill}}, \bibinfo
  {author} {\bibfnamefont {L.}~\bibnamefont {Isenhower}}, \ and\ \bibinfo
  {author} {\bibfnamefont {T.~G.}\ \bibnamefont {Walker}},\ }\bibfield  {title}
  {\bibinfo {title} {{Rydberg state mediated Quantum gates and entanglement of
  pairs of Neutral atoms}},\ }\href {\doibase 10.1088/1742-6596/264/1/012023}
  {\bibfield  {journal} {\bibinfo  {journal} {J. Phys.: Conf. Ser.}\ }\textbf
  {\bibinfo {volume} {264}},\ \bibinfo {pages} {012023} (\bibinfo {year}
  {2011})}\BibitemShut {NoStop}%
\bibitem [{\citenamefont {Levine}\ \emph {et~al.}()\citenamefont {Levine},
  \citenamefont {Keesling}, \citenamefont {Semeghini}, \citenamefont {Omran},
  \citenamefont {Wang}, \citenamefont {Ebadi}, \citenamefont {Bernien},
  \citenamefont {Greiner}, \citenamefont {Vuleti{\'c}}, \citenamefont
  {Pichler},\ and\ \citenamefont {Lukin}}]{Levine2019}%
  \BibitemOpen
  \bibfield  {author} {\bibinfo {author} {\bibfnamefont {H.}~\bibnamefont
  {Levine}}, \bibinfo {author} {\bibfnamefont {A.}~\bibnamefont {Keesling}},
  \bibinfo {author} {\bibfnamefont {G.}~\bibnamefont {Semeghini}}, \bibinfo
  {author} {\bibfnamefont {A.}~\bibnamefont {Omran}}, \bibinfo {author}
  {\bibfnamefont {T.~T.}\ \bibnamefont {Wang}}, \bibinfo {author}
  {\bibfnamefont {S.}~\bibnamefont {Ebadi}}, \bibinfo {author} {\bibfnamefont
  {H.}~\bibnamefont {Bernien}}, \bibinfo {author} {\bibfnamefont
  {M.}~\bibnamefont {Greiner}}, \bibinfo {author} {\bibfnamefont
  {V.}~\bibnamefont {Vuleti{\'c}}}, \bibinfo {author} {\bibfnamefont
  {H.}~\bibnamefont {Pichler}}, \ and\ \bibinfo {author} {\bibfnamefont
  {M.~D.}\ \bibnamefont {Lukin}},\ }\bibfield  {title} {\bibinfo {title}
  {{Parallel implementation of high-fidelity multi-qubit gates with neutral
  atoms}},\ }\href@noop {} {\bibinfo  {journal} {arXiv: 1908.06101}\
  }\BibitemShut {NoStop}%
\bibitem [{\citenamefont {Graham}\ \emph {et~al.}()\citenamefont {Graham},
  \citenamefont {Kwon}, \citenamefont {Grinkemeyer}, \citenamefont {Marra},
  \citenamefont {Jiang}, \citenamefont {Lichtman}, \citenamefont {Sun},
  \citenamefont {Ebert},\ and\ \citenamefont {Saffman}}]{Graham2019}%
  \BibitemOpen
\bibfield  {journal} {  }\bibfield  {author} {\bibinfo {author} {\bibfnamefont
  {T.~M.}\ \bibnamefont {Graham}}, \bibinfo {author} {\bibfnamefont
  {M.}~\bibnamefont {Kwon}}, \bibinfo {author} {\bibfnamefont {B.}~\bibnamefont
  {Grinkemeyer}}, \bibinfo {author} {\bibfnamefont {Z.}~\bibnamefont {Marra}},
  \bibinfo {author} {\bibfnamefont {X.}~\bibnamefont {Jiang}}, \bibinfo
  {author} {\bibfnamefont {M.~T.}\ \bibnamefont {Lichtman}}, \bibinfo {author}
  {\bibfnamefont {Y.}~\bibnamefont {Sun}}, \bibinfo {author} {\bibfnamefont
  {M.}~\bibnamefont {Ebert}}, \ and\ \bibinfo {author} {\bibfnamefont
  {M.}~\bibnamefont {Saffman}},\ }\bibfield  {title} {\bibinfo {title}
  {{Rydberg mediated entanglement in a two-dimensional neutral atom qubit
  array}},\ }\href@noop {} {\bibinfo  {journal} {arXiv: 1908.06103}\
  }\BibitemShut {NoStop}%
\bibitem [{\citenamefont {Ryabtsev}\ \emph {et~al.}(2011)\citenamefont
  {Ryabtsev}, \citenamefont {Beterov}, \citenamefont {Tretyakov}, \citenamefont
  {Entin},\ and\ \citenamefont {Yakshina}}]{Ryabtsev2011}%
  \BibitemOpen
\bibfield  {journal} {  }\bibfield  {author} {\bibinfo {author} {\bibfnamefont
  {I.~I.}\ \bibnamefont {Ryabtsev}}, \bibinfo {author} {\bibfnamefont {I.~I.}\
  \bibnamefont {Beterov}}, \bibinfo {author} {\bibfnamefont {D.~B.}\
  \bibnamefont {Tretyakov}}, \bibinfo {author} {\bibfnamefont {V.~M.}\
  \bibnamefont {Entin}}, \ and\ \bibinfo {author} {\bibfnamefont {E.~A.}\
  \bibnamefont {Yakshina}},\ }\bibfield  {title} {\bibinfo {title} {{Doppler-
  and recoil-free laser excitation of Rydberg states via three-photon
  transitions}},\ }\href {\doibase 10.1103/PhysRevA.84.053409} {\bibfield
  {journal} {\bibinfo  {journal} {Phys. Rev. A}\ }\textbf {\bibinfo {volume}
  {84}},\ \bibinfo {pages} {053409} (\bibinfo {year} {2011})}\BibitemShut
  {NoStop}%
\bibitem [{\citenamefont {Isenhower}\ \emph {et~al.}(2010)\citenamefont
  {Isenhower}, \citenamefont {Urban}, \citenamefont {Zhang}, \citenamefont
  {Gill}, \citenamefont {Henage}, \citenamefont {Johnson}, \citenamefont
  {Walker},\ and\ \citenamefont {Saffman}}]{Isenhower2010}%
  \BibitemOpen
  \bibfield  {author} {\bibinfo {author} {\bibfnamefont {L.}~\bibnamefont
  {Isenhower}}, \bibinfo {author} {\bibfnamefont {E.}~\bibnamefont {Urban}},
  \bibinfo {author} {\bibfnamefont {X.~L.}\ \bibnamefont {Zhang}}, \bibinfo
  {author} {\bibfnamefont {A.~T.}\ \bibnamefont {Gill}}, \bibinfo {author}
  {\bibfnamefont {T.}~\bibnamefont {Henage}}, \bibinfo {author} {\bibfnamefont
  {T.~A.}\ \bibnamefont {Johnson}}, \bibinfo {author} {\bibfnamefont {T.~G.}\
  \bibnamefont {Walker}}, \ and\ \bibinfo {author} {\bibfnamefont
  {M.}~\bibnamefont {Saffman}},\ }\bibfield  {title} {\bibinfo {title}
  {{Demonstration of a Neutral Atom Controlled-NOT Quantum Gate}},\ }\href@noop
  {} {\bibfield  {journal} {\bibinfo  {journal} {Phys. Rev. Lett.}\ }\textbf
  {\bibinfo {volume} {104}},\ \bibinfo {pages} {010503} (\bibinfo {year}
  {2010})}\BibitemShut {NoStop}%
\bibitem [{\citenamefont {Zhang}\ \emph {et~al.}(2010)\citenamefont {Zhang},
  \citenamefont {Isenhower}, \citenamefont {Gill}, \citenamefont {Walker},\
  and\ \citenamefont {Saffman}}]{Zhang2010}%
  \BibitemOpen
  \bibfield  {author} {\bibinfo {author} {\bibfnamefont {X.~L.}\ \bibnamefont
  {Zhang}}, \bibinfo {author} {\bibfnamefont {L.}~\bibnamefont {Isenhower}},
  \bibinfo {author} {\bibfnamefont {A.~T.}\ \bibnamefont {Gill}}, \bibinfo
  {author} {\bibfnamefont {T.~G.}\ \bibnamefont {Walker}}, \ and\ \bibinfo
  {author} {\bibfnamefont {M.}~\bibnamefont {Saffman}},\ }\bibfield  {title}
  {\bibinfo {title} {{Deterministic entanglement of two neutral atoms via
  Rydberg blockade}},\ }\href@noop {} {\bibfield  {journal} {\bibinfo
  {journal} {Phys. Rev. A}\ }\textbf {\bibinfo {volume} {82}},\ \bibinfo
  {pages} {030306(R)} (\bibinfo {year} {2010})}\BibitemShut {NoStop}%
\bibitem [{\citenamefont {Maller}\ \emph {et~al.}(2015)\citenamefont {Maller},
  \citenamefont {Lichtman}, \citenamefont {Xia}, \citenamefont {Sun},
  \citenamefont {Piotrowicz}, \citenamefont {Carr}, \citenamefont {Isenhower},\
  and\ \citenamefont {Saffman}}]{Maller2015}%
  \BibitemOpen
  \bibfield  {author} {\bibinfo {author} {\bibfnamefont {K.~M.}\ \bibnamefont
  {Maller}}, \bibinfo {author} {\bibfnamefont {M.~T.}\ \bibnamefont
  {Lichtman}}, \bibinfo {author} {\bibfnamefont {T.}~\bibnamefont {Xia}},
  \bibinfo {author} {\bibfnamefont {Y.}~\bibnamefont {Sun}}, \bibinfo {author}
  {\bibfnamefont {M.~J.}\ \bibnamefont {Piotrowicz}}, \bibinfo {author}
  {\bibfnamefont {A.~W.}\ \bibnamefont {Carr}}, \bibinfo {author}
  {\bibfnamefont {L.}~\bibnamefont {Isenhower}}, \ and\ \bibinfo {author}
  {\bibfnamefont {M.}~\bibnamefont {Saffman}},\ }\bibfield  {title} {\bibinfo
  {title} {{Rydberg-blockade controlled-not gate and entanglement in a
  two-dimensional array of neutral-atom qubits}},\ }\href@noop {} {\bibfield
  {journal} {\bibinfo  {journal} {Phys. Rev. A}\ }\textbf {\bibinfo {volume}
  {92}},\ \bibinfo {pages} {022336} (\bibinfo {year} {2015})}\BibitemShut
  {NoStop}%
\bibitem [{\citenamefont {Zeng}\ \emph {et~al.}(2017)\citenamefont {Zeng},
  \citenamefont {Xu}, \citenamefont {He}, \citenamefont {Liu}, \citenamefont
  {Liu}, \citenamefont {Wang}, \citenamefont {Papoular}, \citenamefont
  {Shlyapnikov},\ and\ \citenamefont {Zhan}}]{Zeng2017}%
  \BibitemOpen
  \bibfield  {author} {\bibinfo {author} {\bibfnamefont {Y.}~\bibnamefont
  {Zeng}}, \bibinfo {author} {\bibfnamefont {P.}~\bibnamefont {Xu}}, \bibinfo
  {author} {\bibfnamefont {X.}~\bibnamefont {He}}, \bibinfo {author}
  {\bibfnamefont {Y.}~\bibnamefont {Liu}}, \bibinfo {author} {\bibfnamefont
  {M.}~\bibnamefont {Liu}}, \bibinfo {author} {\bibfnamefont {J.}~\bibnamefont
  {Wang}}, \bibinfo {author} {\bibfnamefont {D.~J.}\ \bibnamefont {Papoular}},
  \bibinfo {author} {\bibfnamefont {G.~V.}\ \bibnamefont {Shlyapnikov}}, \ and\
  \bibinfo {author} {\bibfnamefont {M.}~\bibnamefont {Zhan}},\ }\bibfield
  {title} {\bibinfo {title} {{Entangling Two Individual Atoms of Different
  Isotopes via Rydberg Blockade}},\ }\href@noop {} {\bibfield  {journal}
  {\bibinfo  {journal} {Phys. Rev. Lett.}\ }\textbf {\bibinfo {volume} {119}},\
  \bibinfo {pages} {160502} (\bibinfo {year} {2017})}\BibitemShut {NoStop}%
\bibitem [{\citenamefont {Levine}\ \emph {et~al.}(2018)\citenamefont {Levine},
  \citenamefont {Keesling}, \citenamefont {Omran}, \citenamefont {Bernien},
  \citenamefont {Schwartz}, \citenamefont {Zibrov}, \citenamefont {Endres},
  \citenamefont {Greiner}, \citenamefont {Vuleti{\'c}},\ and\ \citenamefont
  {Lukin}}]{Levine2018}%
  \BibitemOpen
  \bibfield  {author} {\bibinfo {author} {\bibfnamefont {H.}~\bibnamefont
  {Levine}}, \bibinfo {author} {\bibfnamefont {A.}~\bibnamefont {Keesling}},
  \bibinfo {author} {\bibfnamefont {A.}~\bibnamefont {Omran}}, \bibinfo
  {author} {\bibfnamefont {H.}~\bibnamefont {Bernien}}, \bibinfo {author}
  {\bibfnamefont {S.}~\bibnamefont {Schwartz}}, \bibinfo {author}
  {\bibfnamefont {A.~S.}\ \bibnamefont {Zibrov}}, \bibinfo {author}
  {\bibfnamefont {M.}~\bibnamefont {Endres}}, \bibinfo {author} {\bibfnamefont
  {M.}~\bibnamefont {Greiner}}, \bibinfo {author} {\bibfnamefont
  {V.}~\bibnamefont {Vuleti{\'c}}}, \ and\ \bibinfo {author} {\bibfnamefont
  {M.~D.}\ \bibnamefont {Lukin}},\ }\bibfield  {title} {\bibinfo {title}
  {{High-fidelity control and entanglement of Rydberg atom qubits}},\ }\href
  {\doibase 10.1103/PhysRevLett.121.123603} {\bibfield  {journal} {\bibinfo
  {journal} {Phys. Rev. Lett.}\ }\textbf {\bibinfo {volume} {121}},\ \bibinfo
  {pages} {123603} (\bibinfo {year} {2018})}\BibitemShut {NoStop}%
\bibitem [{\citenamefont {Picken}\ \emph {et~al.}(2019)\citenamefont {Picken},
  \citenamefont {Legaie}, \citenamefont {McDonnell},\ and\ \citenamefont
  {Pritchard}}]{Picken2018}%
  \BibitemOpen
  \bibfield  {author} {\bibinfo {author} {\bibfnamefont {C.~J.}\ \bibnamefont
  {Picken}}, \bibinfo {author} {\bibfnamefont {R.}~\bibnamefont {Legaie}},
  \bibinfo {author} {\bibfnamefont {K.}~\bibnamefont {McDonnell}}, \ and\
  \bibinfo {author} {\bibfnamefont {J.~D.}\ \bibnamefont {Pritchard}},\
  }\bibfield  {title} {\bibinfo {title} {{Entanglement of neutral-atom qubits
  with long ground-Rydberg coherence times}},\ }\href@noop {} {\bibfield
  {journal} {\bibinfo  {journal} {Quantum Sci. Technol.}\ }\textbf {\bibinfo
  {volume} {4}},\ \bibinfo {pages} {015011} (\bibinfo {year}
  {2019})}\BibitemShut {NoStop}%
\bibitem [{\citenamefont {de~L{\'e}s{\'e}leuc}\ \emph
  {et~al.}(2018)\citenamefont {de~L{\'e}s{\'e}leuc}, \citenamefont {Barredo},
  \citenamefont {Lienhard}, \citenamefont {Browaeys},\ and\ \citenamefont
  {Lahaye}}]{DeLeseleuc2018}%
  \BibitemOpen
  \bibfield  {author} {\bibinfo {author} {\bibfnamefont {S.}~\bibnamefont
  {de~L{\'e}s{\'e}leuc}}, \bibinfo {author} {\bibfnamefont {D.}~\bibnamefont
  {Barredo}}, \bibinfo {author} {\bibfnamefont {V.}~\bibnamefont {Lienhard}},
  \bibinfo {author} {\bibfnamefont {A.}~\bibnamefont {Browaeys}}, \ and\
  \bibinfo {author} {\bibfnamefont {T.}~\bibnamefont {Lahaye}},\ }\bibfield
  {title} {\bibinfo {title} {{Analysis of imperfections in the coherent optical
  excitation of single atoms to Rydberg states}},\ }\href {\doibase
  10.1103/PhysRevA.97.053803} {\bibfield  {journal} {\bibinfo  {journal} {Phys.
  Rev. A}\ }\textbf {\bibinfo {volume} {97}},\ \bibinfo {pages} {053803}
  (\bibinfo {year} {2018})}\BibitemShut {NoStop}%
\bibitem [{\citenamefont {Shi}(2019)}]{Shi2018Accuv1}%
  \BibitemOpen
  \bibfield  {author} {\bibinfo {author} {\bibfnamefont {X.-F.}\ \bibnamefont
  {Shi}},\ }\bibfield  {title} {\bibinfo {title} {{Fast, Accurate, and
  Realizable Two-Qubit Entangling Gates by Quantum Interference in Detuned Rabi
  Cycles of Rydberg Atoms}},\ }\href@noop {} {\bibfield  {journal} {\bibinfo
  {journal} {Phys. Rev. Appl.}\ }\textbf {\bibinfo {volume} {11}},\ \bibinfo
  {pages} {044035} (\bibinfo {year} {2019})}\BibitemShut {NoStop}%
\bibitem [{Shi()}]{Shi2019Dopp2}%
  \BibitemOpen
  \bibfield  {title} {\bibinfo {title} {{Suppress motional dephasing of
  ground-Rydberg transition for high-fidelity quantum control with neutral
  atoms}},\ }\href@noop {} {\bibinfo  {journal} {arXiv: 1909.03857}\
  }\BibitemShut {NoStop}%
\bibitem [{\citenamefont {Shi}\ and\ \citenamefont
  {Kennedy}(2018)}]{Shi2018pra_m}%
  \BibitemOpen
\bibfield  {journal} {  }\bibfield  {author} {\bibinfo {author} {\bibfnamefont
  {X.-F.}\ \bibnamefont {Shi}}\ and\ \bibinfo {author} {\bibfnamefont
  {T.~A.~B.}\ \bibnamefont {Kennedy}},\ }\bibfield  {title} {\bibinfo {title}
  {{Simulating magnetic fields in Rydberg-dressed neutral atoms}},\ }\href
  {\doibase 10.1103/PhysRevA.97.033414} {\bibfield  {journal} {\bibinfo
  {journal} {Phys. Rev. A}\ }\textbf {\bibinfo {volume} {97}},\ \bibinfo
  {pages} {033414} (\bibinfo {year} {2018})}\BibitemShut {NoStop}%
\bibitem [{\citenamefont {Covey}\ \emph {et~al.}(2019)\citenamefont {Covey},
  \citenamefont {Sipahigil}, \citenamefont {Szoke}, \citenamefont {Sinclair},
  \citenamefont {Endres},\ and\ \citenamefont {Painter}}]{Covey2019prappl}%
  \BibitemOpen
  \bibfield  {author} {\bibinfo {author} {\bibfnamefont {J.~P.}\ \bibnamefont
  {Covey}}, \bibinfo {author} {\bibfnamefont {A.}~\bibnamefont {Sipahigil}},
  \bibinfo {author} {\bibfnamefont {S.}~\bibnamefont {Szoke}}, \bibinfo
  {author} {\bibfnamefont {N.}~\bibnamefont {Sinclair}}, \bibinfo {author}
  {\bibfnamefont {M.}~\bibnamefont {Endres}}, \ and\ \bibinfo {author}
  {\bibfnamefont {O.}~\bibnamefont {Painter}},\ }\bibfield  {title} {\bibinfo
  {title} {{Telecom-Band Quantum Optics with Ytterbium Atoms and Silicon
  Nanophotonics}},\ }\href {\doibase 10.1103/PhysRevApplied.11.034044}
  {\bibfield  {journal} {\bibinfo  {journal} {Phys. Rev. Appl.}\ }\textbf
  {\bibinfo {volume} {11}},\ \bibinfo {pages} {034044} (\bibinfo {year}
  {2019})}\BibitemShut {NoStop}%
\bibitem [{\citenamefont {Goreslavsky}\ \emph {et~al.}(1980)\citenamefont
  {Goreslavsky}, \citenamefont {Delone},\ and\ \citenamefont
  {Krainov}}]{Goreslavsky1980}%
  \BibitemOpen
  \bibfield  {author} {\bibinfo {author} {\bibfnamefont {S.~P.}\ \bibnamefont
  {Goreslavsky}}, \bibinfo {author} {\bibfnamefont {N.~B.}\ \bibnamefont
  {Delone}}, \ and\ \bibinfo {author} {\bibfnamefont {V.~P.}\ \bibnamefont
  {Krainov}},\ }\bibfield  {title} {\bibinfo {title} {{The dynamics and
  spontaneous radiation of a two-level atom in a bichromatic field}},\
  }\href@noop {} {\bibfield  {journal} {\bibinfo  {journal} {J. Phys. B}\
  }\textbf {\bibinfo {volume} {13}},\ \bibinfo {pages} {2659} (\bibinfo {year}
  {1980})}\BibitemShut {NoStop}%
\bibitem [{\citenamefont {Baur}\ \emph {et~al.}(2014)\citenamefont {Baur},
  \citenamefont {Tiarks}, \citenamefont {Rempe},\ and\ \citenamefont
  {D{\"u}rr}}]{Baur2014}%
  \BibitemOpen
  \bibfield  {author} {\bibinfo {author} {\bibfnamefont {S.}~\bibnamefont
  {Baur}}, \bibinfo {author} {\bibfnamefont {D.}~\bibnamefont {Tiarks}},
  \bibinfo {author} {\bibfnamefont {G.}~\bibnamefont {Rempe}}, \ and\ \bibinfo
  {author} {\bibfnamefont {S.}~\bibnamefont {D{\"u}rr}},\ }\bibfield  {title}
  {\bibinfo {title} {{Single-Photon Switch Based on Rydberg Blockade}},\
  }\href@noop {} {\bibfield  {journal} {\bibinfo  {journal} {Phys. Rev. Lett.}\
  }\textbf {\bibinfo {volume} {112}},\ \bibinfo {pages} {073901} (\bibinfo
  {year} {2014})}\BibitemShut {NoStop}%
\bibitem [{\citenamefont {Sansonetti}(2006)}]{Sansonetti2006}%
  \BibitemOpen
  \bibfield  {author} {\bibinfo {author} {\bibfnamefont {J.~E.}\ \bibnamefont
  {Sansonetti}},\ }\bibfield  {title} {\bibinfo {title} {{Wavelengths,
  Transition Probabilities, and Energy Levels for the Spectra of Rubidium (Rb I
  through Rb XXXVII)}},\ }\href {\doibase 10.1063/1.2035727} {\bibfield
  {journal} {\bibinfo  {journal} {J. Phys. Chem. Ref. Data}\ }\textbf {\bibinfo
  {volume} {35}},\ \bibinfo {pages} {301} (\bibinfo {year} {2006})}\BibitemShut
  {NoStop}%
\bibitem [{\citenamefont {Sansonetti}(2009)}]{Sansonetti2009}%
  \BibitemOpen
  \bibfield  {author} {\bibinfo {author} {\bibfnamefont {J.~E.}\ \bibnamefont
  {Sansonetti}},\ }\bibfield  {title} {\bibinfo {title} {{Wavelengths,
  Transition Probabilities, and Energy Levels for the Spectra of Cesium Cs I
  --Cs LV}},\ }\href {\doibase 10.1063/1.3132702} {\bibfield  {journal}
  {\bibinfo  {journal} {J. Phys. Chem. Ref. Data}\ }\textbf {\bibinfo {volume}
  {38}},\ \bibinfo {pages} {761} (\bibinfo {year} {2009})}\BibitemShut
  {NoStop}%
\bibitem [{\citenamefont {Pedersen}\ \emph {et~al.}(2007)\citenamefont
  {Pedersen}, \citenamefont {M{\o}ller},\ and\ \citenamefont
  {M{\o}lmer}}]{Pedersen2007}%
  \BibitemOpen
  \bibfield  {author} {\bibinfo {author} {\bibfnamefont {L.~H.}\ \bibnamefont
  {Pedersen}}, \bibinfo {author} {\bibfnamefont {N.~M.}\ \bibnamefont
  {M{\o}ller}}, \ and\ \bibinfo {author} {\bibfnamefont {K.}~\bibnamefont
  {M{\o}lmer}},\ }\bibfield  {title} {\bibinfo {title} {{Fidelity of quantum
  operations}},\ }\href@noop {} {\bibfield  {journal} {\bibinfo  {journal}
  {Phys. Lett. A}\ }\textbf {\bibinfo {volume} {367}},\ \bibinfo {pages} {47}
  (\bibinfo {year} {2007})}\BibitemShut {NoStop}%
\bibitem [{\citenamefont {Zhang}\ \emph {et~al.}(2012)\citenamefont {Zhang},
  \citenamefont {Gill}, \citenamefont {Isenhower}, \citenamefont {Walker},\
  and\ \citenamefont {Saffman}}]{Zhang2012}%
  \BibitemOpen
  \bibfield  {author} {\bibinfo {author} {\bibfnamefont {X.~L.}\ \bibnamefont
  {Zhang}}, \bibinfo {author} {\bibfnamefont {A.~T.}\ \bibnamefont {Gill}},
  \bibinfo {author} {\bibfnamefont {L.}~\bibnamefont {Isenhower}}, \bibinfo
  {author} {\bibfnamefont {T.~G.}\ \bibnamefont {Walker}}, \ and\ \bibinfo
  {author} {\bibfnamefont {M.}~\bibnamefont {Saffman}},\ }\bibfield  {title}
  {\bibinfo {title} {{Fidelity of a Rydberg-blockade quantum gate from
  simulated quantum process tomography}},\ }\href@noop {} {\bibfield  {journal}
  {\bibinfo  {journal} {Phys. Rev. A}\ }\textbf {\bibinfo {volume} {85}},\
  \bibinfo {pages} {042310} (\bibinfo {year} {2012})}\BibitemShut {NoStop}%
\bibitem [{\citenamefont {Li}\ \emph {et~al.}(2003)\citenamefont {Li},
  \citenamefont {Mourachko}, \citenamefont {Noel},\ and\ \citenamefont
  {Gallagher}}]{Li2003}%
  \BibitemOpen
  \bibfield  {author} {\bibinfo {author} {\bibfnamefont {W.}~\bibnamefont
  {Li}}, \bibinfo {author} {\bibfnamefont {I.}~\bibnamefont {Mourachko}},
  \bibinfo {author} {\bibfnamefont {M.~W.}\ \bibnamefont {Noel}}, \ and\
  \bibinfo {author} {\bibfnamefont {T.~F.}\ \bibnamefont {Gallagher}},\
  }\bibfield  {title} {\bibinfo {title} {{Millimeter-wave spectroscopy of cold
  Rb Rydberg atoms in a magneto-optical trap: Quantum defects of the ns, np,
  and nd series}},\ }\href@noop {} {\bibfield  {journal} {\bibinfo  {journal}
  {Phys. Rev. A}\ }\textbf {\bibinfo {volume} {67}},\ \bibinfo {pages} {052502}
  (\bibinfo {year} {2003})}\BibitemShut {NoStop}%
\bibitem [{\citenamefont {Han}\ \emph {et~al.}(2006)\citenamefont {Han},
  \citenamefont {Jamil}, \citenamefont {Norum}, \citenamefont {Tanner},\ and\
  \citenamefont {Gallagher}}]{Han2006}%
  \BibitemOpen
  \bibfield  {author} {\bibinfo {author} {\bibfnamefont {J.}~\bibnamefont
  {Han}}, \bibinfo {author} {\bibfnamefont {Y.}~\bibnamefont {Jamil}}, \bibinfo
  {author} {\bibfnamefont {D.~V.~L.}\ \bibnamefont {Norum}}, \bibinfo {author}
  {\bibfnamefont {P.~J.}\ \bibnamefont {Tanner}}, \ and\ \bibinfo {author}
  {\bibfnamefont {T.~F.}\ \bibnamefont {Gallagher}},\ }\bibfield  {title}
  {\bibinfo {title} {{Rb nf quantum defects from millimeter-wave spectroscopy
  of cold $^{85}$Rb Rydberg atoms}},\ }\href {\doibase
  10.1103/PhysRevA.74.054502} {\bibfield  {journal} {\bibinfo  {journal} {Phys.
  Rev. A}\ }\textbf {\bibinfo {volume} {74}},\ \bibinfo {pages} {054502}
  (\bibinfo {year} {2006})}\BibitemShut {NoStop}%
\bibitem [{\citenamefont {Walker}\ and\ \citenamefont
  {Saffman}(2008)}]{Walker2008}%
  \BibitemOpen
  \bibfield  {author} {\bibinfo {author} {\bibfnamefont {T.~G.}\ \bibnamefont
  {Walker}}\ and\ \bibinfo {author} {\bibfnamefont {M.}~\bibnamefont
  {Saffman}},\ }\bibfield  {title} {\bibinfo {title} {{Consequences of Zeeman
  degeneracy for the van der Waals blockade between Rydberg atoms}},\
  }\href@noop {} {\bibfield  {journal} {\bibinfo  {journal} {Phys. Rev. A}\
  }\textbf {\bibinfo {volume} {77}},\ \bibinfo {pages} {032723} (\bibinfo
  {year} {2008})}\BibitemShut {NoStop}%
\bibitem [{\citenamefont {Beterov}\ \emph {et~al.}(2009)\citenamefont
  {Beterov}, \citenamefont {Ryabtsev}, \citenamefont {Tretyakov},\ and\
  \citenamefont {Entin}}]{Beterov2009}%
  \BibitemOpen
  \bibfield  {author} {\bibinfo {author} {\bibfnamefont {I.~I.}\ \bibnamefont
  {Beterov}}, \bibinfo {author} {\bibfnamefont {I.~I.}\ \bibnamefont
  {Ryabtsev}}, \bibinfo {author} {\bibfnamefont {D.~B.}\ \bibnamefont
  {Tretyakov}}, \ and\ \bibinfo {author} {\bibfnamefont {V.~M.}\ \bibnamefont
  {Entin}},\ }\bibfield  {title} {\bibinfo {title} {{Quasiclassical
  calculations of blackbody-radiation-induced depopulation rates and effective
  lifetimes of Rydberg nS, nP, and nD alkali-metal atoms with n$\leq$80}},\
  }\href {\doibase 10.1103/PhysRevA.79.052504} {\bibfield  {journal} {\bibinfo
  {journal} {Phys. Rev. A}\ }\textbf {\bibinfo {volume} {79}},\ \bibinfo
  {pages} {052504} (\bibinfo {year} {2009})}\BibitemShut {NoStop}%
\bibitem [{\citenamefont {Saffman}\ and\ \citenamefont
  {Walker}(2005)}]{Saffman2005}%
  \BibitemOpen
  \bibfield  {author} {\bibinfo {author} {\bibfnamefont {M.}~\bibnamefont
  {Saffman}}\ and\ \bibinfo {author} {\bibfnamefont {T.~G.}\ \bibnamefont
  {Walker}},\ }\bibfield  {title} {\bibinfo {title} {{Analysis of a quantum
  logic device based on dipole-dipole interactions of optically trapped Rydberg
  atoms}},\ }\href@noop {} {\bibfield  {journal} {\bibinfo  {journal} {Phys.
  Rev. A}\ }\textbf {\bibinfo {volume} {72}},\ \bibinfo {pages} {022347}
  (\bibinfo {year} {2005})}\BibitemShut {NoStop}%
\bibitem [{\citenamefont {Crow}\ \emph {et~al.}(2016)\citenamefont {Crow},
  \citenamefont {Joynt},\ and\ \citenamefont {Saffman}}]{Crow2016}%
  \BibitemOpen
  \bibfield  {author} {\bibinfo {author} {\bibfnamefont {D.}~\bibnamefont
  {Crow}}, \bibinfo {author} {\bibfnamefont {R.}~\bibnamefont {Joynt}}, \ and\
  \bibinfo {author} {\bibfnamefont {M.}~\bibnamefont {Saffman}},\ }\bibfield
  {title} {\bibinfo {title} {{Improved Error Thresholds for Measurement-Free
  Error Correction}},\ }\href {\doibase 10.1103/PhysRevLett.117.130503}
  {\bibfield  {journal} {\bibinfo  {journal} {Phys. Rev. Lett.}\ }\textbf
  {\bibinfo {volume} {117}},\ \bibinfo {pages} {130503} (\bibinfo {year}
  {2016})}\BibitemShut {NoStop}%
\bibitem [{\citenamefont {Shi}(2018{\natexlab{a}})}]{Shi2018prapp2}%
  \BibitemOpen
  \bibfield  {author} {\bibinfo {author} {\bibfnamefont {X.-F.}\ \bibnamefont
  {Shi}},\ }\bibfield  {title} {\bibinfo {title} {{Accurate Quantum Logic Gates
  by Spin Echo in Rydberg Atoms}},\ }\href {\doibase
  10.1103/PhysRevApplied.7.064017} {\bibfield  {journal} {\bibinfo  {journal}
  {Phys. Rev. Appl.}\ }\textbf {\bibinfo {volume} {10}},\ \bibinfo {pages}
  {034006} (\bibinfo {year} {2018}{\natexlab{a}})}\BibitemShut {NoStop}%
\bibitem [{\citenamefont {Shi}\ \emph {et~al.}(2016)\citenamefont {Shi},
  \citenamefont {Svetlichnyy},\ and\ \citenamefont {Kennedy}}]{ShiJPB2016}%
  \BibitemOpen
  \bibfield  {author} {\bibinfo {author} {\bibfnamefont {X.-F.}\ \bibnamefont
  {Shi}}, \bibinfo {author} {\bibfnamefont {P.}~\bibnamefont {Svetlichnyy}}, \
  and\ \bibinfo {author} {\bibfnamefont {T.~A.~B.}\ \bibnamefont {Kennedy}},\
  }\bibfield  {title} {\bibinfo {title} {{Spin -- charge separation of
  dark-state polaritons in a Rydberg medium}},\ }\href@noop {} {\bibfield
  {journal} {\bibinfo  {journal} {J. Phys. B}\ }\textbf {\bibinfo {volume}
  {49}},\ \bibinfo {pages} {074005} (\bibinfo {year} {2016})}\BibitemShut
  {NoStop}%
\bibitem [{\citenamefont {Shi}\ and\ \citenamefont
  {Kennedy}(2017)}]{Shi2017pra}%
  \BibitemOpen
  \bibfield  {author} {\bibinfo {author} {\bibfnamefont {X.-F.}\ \bibnamefont
  {Shi}}\ and\ \bibinfo {author} {\bibfnamefont {T.~A.~B.}\ \bibnamefont
  {Kennedy}},\ }\bibfield  {title} {\bibinfo {title} {{Annulled van der Waals
  interaction and fast Rydberg quantum gates}},\ }\href {\doibase
  10.1103/PhysRevA.95.043429} {\bibfield  {journal} {\bibinfo  {journal} {Phys.
  Rev. A}\ }\textbf {\bibinfo {volume} {95}},\ \bibinfo {pages} {043429}
  (\bibinfo {year} {2017})}\BibitemShut {NoStop}%
\bibitem [{\citenamefont {Shi}(2018{\natexlab{b}})}]{Shi2018a}%
  \BibitemOpen
  \bibfield  {author} {\bibinfo {author} {\bibfnamefont {X.-F.}\ \bibnamefont
  {Shi}},\ }\bibfield  {title} {\bibinfo {title} {{Universal Barenco quantum
  gates via a tunable non-collinear interaction}},\ }\href {\doibase
  10.1103/PhysRevA.97.032310} {\bibfield  {journal} {\bibinfo  {journal} {Phys.
  Rev. A}\ }\textbf {\bibinfo {volume} {97}},\ \bibinfo {pages} {032310}
  (\bibinfo {year} {2018}{\natexlab{b}})}\BibitemShut {NoStop}%
\bibitem [{\citenamefont {Shi}(2017)}]{Shi2017}%
  \BibitemOpen
  \bibfield  {author} {\bibinfo {author} {\bibfnamefont {X.-F.}\ \bibnamefont
  {Shi}},\ }\bibfield  {title} {\bibinfo {title} {{Rydberg Quantum Gates Free
  from Blockade Error}},\ }\href {\doibase 10.1103/PhysRevApplied.7.064017}
  {\bibfield  {journal} {\bibinfo  {journal} {Phys. Rev. Appl.}\ }\textbf
  {\bibinfo {volume} {7}},\ \bibinfo {pages} {064017} (\bibinfo {year}
  {2017})}\BibitemShut {NoStop}%
\bibitem [{\citenamefont {Goerz}\ \emph {et~al.}(2014)\citenamefont {Goerz},
  \citenamefont {Halperin}, \citenamefont {Aytac}, \citenamefont {Koch},\ and\
  \citenamefont {Whaley}}]{Goerz2014}%
  \BibitemOpen
  \bibfield  {author} {\bibinfo {author} {\bibfnamefont {M.~H.}\ \bibnamefont
  {Goerz}}, \bibinfo {author} {\bibfnamefont {E.~J.}\ \bibnamefont {Halperin}},
  \bibinfo {author} {\bibfnamefont {J.~M.}\ \bibnamefont {Aytac}}, \bibinfo
  {author} {\bibfnamefont {C.~P.}\ \bibnamefont {Koch}}, \ and\ \bibinfo
  {author} {\bibfnamefont {K.~B.}\ \bibnamefont {Whaley}},\ }\bibfield  {title}
  {\bibinfo {title} {{Robustness of high-fidelity Rydberg gates with
  single-site addressability}},\ }\href@noop {} {\bibfield  {journal} {\bibinfo
   {journal} {Phys. Rev. A}\ }\textbf {\bibinfo {volume} {90}},\ \bibinfo
  {pages} {032329} (\bibinfo {year} {2014})}\BibitemShut {NoStop}%
\bibitem [{\citenamefont {Theis}\ \emph {et~al.}(2016)\citenamefont {Theis},
  \citenamefont {Motzoi}, \citenamefont {Wilhelm},\ and\ \citenamefont
  {Saffman}}]{Theis2016}%
  \BibitemOpen
  \bibfield  {author} {\bibinfo {author} {\bibfnamefont {L.~S.}\ \bibnamefont
  {Theis}}, \bibinfo {author} {\bibfnamefont {F.}~\bibnamefont {Motzoi}},
  \bibinfo {author} {\bibfnamefont {F.~K.}\ \bibnamefont {Wilhelm}}, \ and\
  \bibinfo {author} {\bibfnamefont {M.}~\bibnamefont {Saffman}},\ }\bibfield
  {title} {\bibinfo {title} {{High-fidelity Rydberg-blockade entangling gate
  using shaped, analytic pulses}},\ }\href@noop {} {\bibfield  {journal}
  {\bibinfo  {journal} {Phys. Rev. A}\ }\textbf {\bibinfo {volume} {94}},\
  \bibinfo {pages} {032306} (\bibinfo {year} {2016})}\BibitemShut {NoStop}%
\bibitem [{\citenamefont {Petrosyan}\ \emph {et~al.}(2017)\citenamefont
  {Petrosyan}, \citenamefont {Motzoi}, \citenamefont {Saffman},\ and\
  \citenamefont {M{\o}lmer}}]{Petrosyan2017}%
  \BibitemOpen
  \bibfield  {author} {\bibinfo {author} {\bibfnamefont {D.}~\bibnamefont
  {Petrosyan}}, \bibinfo {author} {\bibfnamefont {F.}~\bibnamefont {Motzoi}},
  \bibinfo {author} {\bibfnamefont {M.}~\bibnamefont {Saffman}}, \ and\
  \bibinfo {author} {\bibfnamefont {K.}~\bibnamefont {M{\o}lmer}},\ }\bibfield
  {title} {\bibinfo {title} {{High-fidelity Rydberg quantum gate via a two-atom
  dark state}},\ }\href@noop {} {\bibfield  {journal} {\bibinfo  {journal}
  {Phys. Rev. A}\ }\textbf {\bibinfo {volume} {96}},\ \bibinfo {pages} {042306}
  (\bibinfo {year} {2017})}\BibitemShut {NoStop}%
\bibitem [{\citenamefont {Yu}\ \emph {et~al.}(2019)\citenamefont {Yu},
  \citenamefont {Wang}, \citenamefont {Ma}, \citenamefont {Zhao},\ and\
  \citenamefont {Qian}}]{Yu2019}%
  \BibitemOpen
  \bibfield  {author} {\bibinfo {author} {\bibfnamefont {D.}~\bibnamefont
  {Yu}}, \bibinfo {author} {\bibfnamefont {H.}~\bibnamefont {Wang}}, \bibinfo
  {author} {\bibfnamefont {D.}~\bibnamefont {Ma}}, \bibinfo {author}
  {\bibfnamefont {X.-D.}\ \bibnamefont {Zhao}}, \ and\ \bibinfo {author}
  {\bibfnamefont {J.}~\bibnamefont {Qian}},\ }\bibfield  {title} {\bibinfo
  {title} {{Adiabatic and high-fidelity quantum gates with hybrid
  Rydberg-Rydberg interactions}},\ }\href {http://arxiv.org/abs/1905.10937}
  {\bibfield  {journal} {\bibinfo  {journal} {Opt. Express}\ }\textbf {\bibinfo
  {volume} {27}},\ \bibinfo {pages} {23080} (\bibinfo {year}
  {2019})}\BibitemShut {NoStop}%
\bibitem [{\citenamefont {Shen}\ \emph {et~al.}(2019)\citenamefont {Shen},
  \citenamefont {Wu}, \citenamefont {Su},\ and\ \citenamefont
  {Liang}}]{Shen:19}%
  \BibitemOpen
  \bibfield  {author} {\bibinfo {author} {\bibfnamefont {C.-P.}\ \bibnamefont
  {Shen}}, \bibinfo {author} {\bibfnamefont {J.-L.}\ \bibnamefont {Wu}},
  \bibinfo {author} {\bibfnamefont {S.-L.}\ \bibnamefont {Su}}, \ and\ \bibinfo
  {author} {\bibfnamefont {E.}~\bibnamefont {Liang}},\ }\bibfield  {title}
  {\bibinfo {title} {{Construction of robust Rydberg controlled-phase gates}},\
  }\href {\doibase 10.1364/OL.44.002036} {\bibfield  {journal} {\bibinfo
  {journal} {Opt. Lett.}\ }\textbf {\bibinfo {volume} {44}},\ \bibinfo {pages}
  {2036} (\bibinfo {year} {2019})}\BibitemShut {NoStop}%
\bibitem [{\citenamefont {Liao}\ \emph {et~al.}(2019)\citenamefont {Liao},
  \citenamefont {Lu}, \citenamefont {Li},\ and\ \citenamefont {Du}}]{Liao2019}%
  \BibitemOpen
  \bibfield  {author} {\bibinfo {author} {\bibfnamefont {K.-Y.}\ \bibnamefont
  {Liao}}, \bibinfo {author} {\bibfnamefont {X.-H.}\ \bibnamefont {Lu}},
  \bibinfo {author} {\bibfnamefont {Z.}~\bibnamefont {Li}}, \ and\ \bibinfo
  {author} {\bibfnamefont {Y.-X.}\ \bibnamefont {Du}},\ }\bibfield  {title}
  {\bibinfo {title} {{Geometric Rydberg quantum gate with shortcuts to
  adiabaticity}},\ }\href@noop {} {\bibfield  {journal} {\bibinfo  {journal}
  {Opt. Lett.}\ }\textbf {\bibinfo {volume} {44}},\ \bibinfo {pages} {4801}
  (\bibinfo {year} {2019})}\BibitemShut {NoStop}%
\bibitem [{\citenamefont {Isenhower}\ \emph {et~al.}(2011)\citenamefont
  {Isenhower}, \citenamefont {Saffman},\ and\ \citenamefont
  {M{\o}lmer}}]{Isenhower2011}%
  \BibitemOpen
  \bibfield  {author} {\bibinfo {author} {\bibfnamefont {L.}~\bibnamefont
  {Isenhower}}, \bibinfo {author} {\bibfnamefont {M.}~\bibnamefont {Saffman}},
  \ and\ \bibinfo {author} {\bibfnamefont {K.}~\bibnamefont {M{\o}lmer}},\
  }\bibfield  {title} {\bibinfo {title} {{Multibit $C_k$NOT quantum gates via
  Rydberg blockade}},\ }\href {\doibase 10.1007/s11128-011-0292-4} {\bibfield
  {journal} {\bibinfo  {journal} {Quant. Inf. Proc.}\ }\textbf {\bibinfo
  {volume} {10}},\ \bibinfo {pages} {755} (\bibinfo {year} {2011})}\BibitemShut
  {NoStop}%
\bibitem [{\citenamefont {Shi}(2018{\natexlab{c}})}]{Shi2018prapp}%
  \BibitemOpen
  \bibfield  {author} {\bibinfo {author} {\bibfnamefont {X.-F.}\ \bibnamefont
  {Shi}},\ }\bibfield  {title} {\bibinfo {title} {{Deutsch, Toffoli, and CNOT
  Gates via Rydberg Blockade of Neutral Atoms}},\ }\href {\doibase
  10.1103/PhysRevApplied.9.051001} {\bibfield  {journal} {\bibinfo  {journal}
  {Phys. Rev. Appl.}\ }\textbf {\bibinfo {volume} {9}},\ \bibinfo {pages}
  {051001} (\bibinfo {year} {2018}{\natexlab{c}})}\BibitemShut {NoStop}%
\bibitem [{\citenamefont {Li}\ and\ \citenamefont {Shao}(2018)}]{Li2018}%
  \BibitemOpen
  \bibfield  {author} {\bibinfo {author} {\bibfnamefont {D.~X.}\ \bibnamefont
  {Li}}\ and\ \bibinfo {author} {\bibfnamefont {X.~Q.}\ \bibnamefont {Shao}},\
  }\bibfield  {title} {\bibinfo {title} {{Unconventional Rydberg pumping and
  applications in quantum information processing}},\ }\href@noop {} {\bibfield
  {journal} {\bibinfo  {journal} {Phys. Rev. A}\ }\textbf {\bibinfo {volume}
  {98}},\ \bibinfo {pages} {062338} (\bibinfo {year} {2018})}\BibitemShut
  {NoStop}%
\bibitem [{\citenamefont {Beterov}\ \emph {et~al.}(2018)\citenamefont
  {Beterov}, \citenamefont {Ashkarin}, \citenamefont {Yakshina}, \citenamefont
  {Tretyakov}, \citenamefont {Entin}, \citenamefont {Ryabtsev}, \citenamefont
  {Cheinet}, \citenamefont {Pillet},\ and\ \citenamefont
  {Saffman}}]{Beterov2018arX}%
  \BibitemOpen
  \bibfield  {author} {\bibinfo {author} {\bibfnamefont {I.~I.}\ \bibnamefont
  {Beterov}}, \bibinfo {author} {\bibfnamefont {I.~N.}\ \bibnamefont
  {Ashkarin}}, \bibinfo {author} {\bibfnamefont {E.~A.}\ \bibnamefont
  {Yakshina}}, \bibinfo {author} {\bibfnamefont {D.~B.}\ \bibnamefont
  {Tretyakov}}, \bibinfo {author} {\bibfnamefont {V.~M.}\ \bibnamefont
  {Entin}}, \bibinfo {author} {\bibfnamefont {I.~I.}\ \bibnamefont {Ryabtsev}},
  \bibinfo {author} {\bibfnamefont {P.}~\bibnamefont {Cheinet}}, \bibinfo
  {author} {\bibfnamefont {P.}~\bibnamefont {Pillet}}, \ and\ \bibinfo {author}
  {\bibfnamefont {M.}~\bibnamefont {Saffman}},\ }\bibfield  {title} {\bibinfo
  {title} {{Fast three-qubit Toffoli quantum gate based on the three-body
  Forster resonances in Rydberg atoms}},\ }\href
  {http://arxiv.org/abs/1808.03473} {\bibfield  {journal} {\bibinfo  {journal}
  {Phys. Rev. A}\ }\textbf {\bibinfo {volume} {98}},\ \bibinfo {pages} {042704}
  (\bibinfo {year} {2018})}\BibitemShut {NoStop}%
\bibitem [{\citenamefont {Su}\ \emph {et~al.}(2018)\citenamefont {Su},
  \citenamefont {Shen}, \citenamefont {Liang},\ and\ \citenamefont
  {Zhang}}]{Su2018}%
  \BibitemOpen
  \bibfield  {author} {\bibinfo {author} {\bibfnamefont {S.~L.}\ \bibnamefont
  {Su}}, \bibinfo {author} {\bibfnamefont {H.~Z.}\ \bibnamefont {Shen}},
  \bibinfo {author} {\bibfnamefont {E.}~\bibnamefont {Liang}}, \ and\ \bibinfo
  {author} {\bibfnamefont {S.}~\bibnamefont {Zhang}},\ }\bibfield  {title}
  {\bibinfo {title} {{One-step construction of the multiple-qubit Rydberg
  controlled- phase gate}},\ }\href {\doibase 10.1103/PhysRevA.98.032306}
  {\bibfield  {journal} {\bibinfo  {journal} {Phys. Rev. A}\ }\textbf {\bibinfo
  {volume} {98}},\ \bibinfo {pages} {032306} (\bibinfo {year}
  {2018})}\BibitemShut {NoStop}%
\end{thebibliography}
\end{document}